\documentclass[11pt, a4paper]{article}
\usepackage{amsrefs, amsmath, amsthm}
\usepackage{amsfonts, mathrsfs}
\usepackage{amssymb}
\usepackage{amscd, mathtools}
\usepackage{hyperref, eqnarray}
\usepackage{cleveref, multirow, bigdelim} 
\usepackage{fullpage, color}
\usepackage[labelformat=empty]{subfig}
\usepackage{booktabs}

\usepackage{xypic}
\usepackage{dynkin-diagrams}
\usepackage{enumerate, longtable}
\usepackage{makecell, diagbox}
\usepackage{IEEEtrantools}
\usepackage{pifont}
\usepackage{enumitem}
\usepackage{empheq,tabularx}
\usepackage[utf8]{inputenc}

\theoremstyle{plain}

\newtheorem{thm}{Theorem}[section]

\newtheorem{claim}[thm]{Claim}
\newtheorem*{conj*}{Conjecture}

\newtheorem*{cor*}{Corollary}

\theoremstyle{definition}

\newtheorem{example}{Example}

\BibSpec{article}{%
+{}{\sc \PrintAuthors} {author}
+{,}{ \textrm} {title}
+{,}{ \textit} {journal}
+{,}{ } {volume}
+{}{ \IfEmptyBibField{journal}{}{\PrintDate{year}}} {transition}
+{,}{ } {pages}
+{,}{ } {status}
+{,}{ available at \tt} {eprint}
+{.}{} {transition}
}

\BibSpec{book}{%
+{}{\sc \PrintAuthors} {author}
+{,}{ \textrm} {title}
+{,}{ \textit} {series}
+{,}{ \textit} {volume}
+{,}{ \textrm} {publisher}
+{}{ \IfEmptyBibField{publisher}{}{\PrintDate{year}}} {transition}
+{,}{ } {pages}
+{,}{ available at \tt} {eprint}
+{.}{} {transition}
}

\BibSpec{inproceedings}{%
+{}{\sc \PrintAuthors} {author}
+{,}{ \textrm} {title}
+{,}{ \textit} {series}
+{,}{ \textbf} {volume}
+{,}{ \textrm} {publisher}
+{}{ \IfEmptyBibField{publisher}{}{\PrintDate{year}}} {transition}
+{,}{ } {pages}
+{,}{ available at \tt} {eprint}
+{.}{} {transition}
}

\newcommand*\widefbox[1]{\color{red} \fbox{\hspace{2em}#1\hspace{2em}}}

\makeatletter

\crefname{equation}{Eq.}{Eqs.}
\crefname{eqnarray}{Eq.}{Eqs.}
\crefname{algo}{Algorithm}{Algorithms}
\crefname{conj}{Conjecture}{Conjectures}
\crefname{lem}{Lemma}{Lemmas}
\crefname{thm}{Theorem}{Theorems}
\crefname{claim}{Claim}{Claims}
\crefname{rmk}{Remark}{Remarks}
\crefname{prop}{Proposition}{Propositions}
\crefname{section}{Section}{Sections}
\crefname{appendix}{Appendix}{Appendices}
\crefname{cor}{Corollary}{Corollaries}
\crefname{figure}{Figure}{Figures}
\crefname{table}{Table}{Tables}
\crefname{example}{Example}{Examples}
\crefname{prob}{Problem}{Problems}
\crefname{assm}{Assumption}{Assumptions}
\crefname{defn}{Definition}{Definitions}

\newcommand{\Tr}{\mathrm{Tr}\,}
\newcommand{\beq}{\begin{equation}}
\newcommand{\eeq}{\end{equation}}
\newcommand{\bea}{\begin{eqnarray}}
\newcommand{\eea}{\end{eqnarray}}

\newcommand{\bra}{\left\langle}
\newcommand{\ket}{\right\rangle}

\newcommand{\re}{{\rm e}}

\newcommand{\de}{{\partial}}

\newcommand{\rd}{\mathrm{d}}
\newcommand{\ri}{\mathrm{i}}

\newcommand{\bbS}{\mathbb{S}}
\newcommand{\bbN}{\mathbb{N}}

\newcommand{\bbZ}{\mathbb{Z}}
\newcommand{\bbR}{\mathbb{R}}
\newcommand{\bbC}{\mathbb{C}}

\newcommand{\bbP}{\mathbb{P}}
\newcommand{\bbF}{\mathbb{F}}
\newcommand{\bbQ}{\mathbb{Q}}

\def\bary{\begin{array}} 
\def\eary{\end{array}} 
\def\ben{\begin{enumerate}} 
\def\een{\end{enumerate}}
\def\bit{\begin{itemize}} 
\def\eit{\end{itemize}}
\def\nn{\nonumber} 

\newcommand{\cZ}{\mathcal{Z}}
\newcommand{\cO}{\mathcal{O}}

\newcommand{\cC}{\mathcal{C}}
\newcommand{\DD}{\mathcal{D}}

\newcommand{\cK}{\mathcal{K}}
\newcommand{\cN}{\mathcal{N}}
\newcommand{\cW}{\mathcal{W}}
\newcommand{\cG}{\mathcal{G}}
\newcommand{\cA}{\mathcal{A}}

\newcommand{\cF}{\mathcal{F}}
\newcommand{\cI}{\mathcal{I}}

\newcommand{\RR}{\mathcal{R}}

\newcommand{\cV}{\mathcal{V}}

\newcommand{\cM}{\mathcal M}

\begin{document}


\pagestyle{plain}
\addtolength{\baselineskip}{0.10\baselineskip}

\begin{center}

\vspace{26pt}

{\Large \textbf{Five-dimensional gauge theories and the local B-model}}

\vspace{26pt}

\textsl{Andrea Brini$^{1,2}$ and Kento Osuga$^{1,3}$}

\vspace{18pt}

\textsl{$^1$ School of Mathematics and Statistics, 
University of Sheffield, Hounsfield Road, Sheffield S3~7RH, United Kingdom. \\
$^2$ On leave from CNRS, DR13, Montpellier, France. \\
$^3$ Faculty of Physics, University of Warsaw, ul. Pasteura 5, 02-093, Warsaw, Poland.\\}

\vspace{10pt}

E-mail: \texttt{a.brini, k.osuga @sheffield.ac.uk} 
\end{center}

\vspace{20pt}

\begin{center}
\textbf{Abstract}
\end{center}

We propose an effective  framework for computing the prepotential of the topological B-model on a class of local Calabi--Yau geometries related to the circle compactification of five-dimensional $\cN=1$ super Yang--Mills theory with simple gauge group. In the simply-laced case,
we construct Picard--Fuchs operators from 
the Dubrovin connection on the Frobenius manifolds associated to the extended affine Weyl groups of type $\mathrm{ADE}$. In general, we propose a purely algebraic construction of Picard--Fuchs ideals
from a canonical subring of the space of regular functions on the ramification locus of the Seiberg--Witten curve, encompassing non-simply-laced cases as well. \\
We offer several precision tests of our proposal. Whenever a candidate spectral curve is known from string theory/brane engineering,
we perform non-perturbative comparisons with the gauge theory prepotentials obtained from the K-theoretic blow-up equations, finding perfect agreement. 
We also employ our formalism to rule out some proposals from the theory of integrable systems of Seiberg--Witten geometries for non-simply-laced gauge groups.

\noindent

\tableofcontents

\section{Introduction}
\label{sec:intro}

Five-dimensional supersymmetric gauge theories have been the subject of long-standing interest, in light of their intertwined roles both as low-energy descriptions of SCFTs in 5d \cite{Seiberg:1996bd}, and as non-trivial quantum field theories engineered by M-theory compactification on a Calabi--Yau threefold \cite{Intriligator:1997pq,Lawrence:1997jr}. In particular, their stringy origin situates their Kaluza--Klein (KK) reduction on $\bbR^{4}\times S^1$
at the centre of a web of correspondences relating instanton counting to, {\it inter alia}, the topological A-model on resolutions of local CY3 singularities \cite{Katz:1996fh,Lawrence:1997jr}, a class of relativistic integrable models \cite{Nekrasov:1996cz,MR1090424,Fock:2014ifa},  a $q$-deformed version of the AGT correspondence \cite{Awata:2009ur,Nieri:2013yra,Nieri:2013vba}, and in special instances, the large~$N$ limit of Chern--Simons theory on non-trivial 3-manifolds and related matrix models \cite{Aganagic:2002wv,Borot:2014kda,Borot:2015fxa,Marino:2002fk}.
\\

As for any four-dimensional theory with eight supercharges, the infrared physics on the Coulomb branch $\cM_{\mathsf{C}}$ of the $\cN=2$ KK theory on $\bbR^4 \times S^1$ is encoded into its prepotential, which governs the exact Wilsonian effective action of the gauge theory. At weak coupling and for classical gauge groups, one approach to compute this microscopically in the full $\Omega$-background is by using localisation, asymptotically in the Coulomb moduli \cite{Nekrasov:2002qd, Nekrasov:2003rj, Nakajima:2003pg, Nakajima:2005fg}. Alternatively, and equivalently, the $\Omega$-background prepotential coincides with the free energy of the refined A-model on the associated engineering CY3 geometry \cite{Iqbal:2007ii,Awata:2005fa,Awata:2008ed}. 

The spectacular results arising from direct instanton calculations come at a price, however, as explicit instanton partition functions become unwieldy in general, the treatment of exceptional gauge groups requires some degree of guesswork and/or extrapolation from the classical cases, and one is {\it a fortiori} stuck in the S-duality frame corresponding to the instanton expansion/large volume limit of the engineering geometry. By the work of Seiberg and Witten, one way around this is to consider the realisation of the gauge theory prepotential from the special geometry 
of a family of spectral curves fibred
over the Coulomb branch $\cM_\mathsf{C}$ \cite{Seiberg:1994rs}. Restricting for simplicity to the setting of the pure gauge theory with gauge group $\cG$, the affine part $\cC_{\cG;u}=\{\mathsf{P}_{\cG;u}(\mu,\lambda;u)=0\}$
of the fibres of the family over a Coulomb moduli point $u\in\cM_\mathsf{C}$ is given by the vanishing of a certain characteristic polynomial 
$\mathsf{P}_{\cG;u}(\mu,\lambda) \in \bbC[\mu,\lambda]$, and the gauge theory effective action is recovered from the special geometry relations
\beq
\frac{\de \cF_\cG}{\de a_i} = \frac{1}{2} \oint_{B_i} \log \mu ~\rd \log \lambda, \qquad a_i = \frac{1}{2\pi \ri }\oint_{A_i} \log \mu ~\rd \log \lambda,
\label{eq:specgeom}
\eeq
where $\{A_i, B_i\}_{i=1}^r$
is a duality-frame-dependent choice of a symplectic basis of integral homology 1-cycles on $\cC_{\cG;u}$. \medskip

In principle, knowing $\mathsf{P}_{\cG;u}$ determines the prepotential of the gauge theory via the period integrals in \eqref{eq:specgeom}, but translating this into practice, especially when $\cG\neq\mathrm{SU}(N)$, is a much different story. Even being able to write down the integrals (let alone compute them) in \eqref{eq:specgeom} requires a delicate analysis in the choice of a symmetric combination of homology 1-cycles $\{A_i, B_i\}_{i=1}^r$, which realises a projection to a distinguised Prym--Tyurin subvariety\footnote{We denote $\overline{\cC_{\cG;u}}$ the smooth completion (normalisation of the Zariski closure in $\bbP^2$) of the affine curve $\cC_{\cG; u}$. In particular $\overline{\cC_{\cG;u}}$ is a compact Riemann surface.} of $\mathrm{Jac}(\overline{\cC_{\cG;u}})$, see \cite{Martinec:1995by, Brini:2017gfi}. On top of this, the curves $\overline{\cC_{\cG;u}}$ are usually non-hyperelliptic and of high genus on a dense open subset of 
$\cM_\mathsf{C}$, {\it de facto} hampering a direct calculation of the period integrals. 
One possible perspective to overcome this impasse is offered by string theory engineering, wherein for $\cG=\mathrm{SU}(N)$ the SW geometry gets identified with the Hori--Iqbal--Vafa mirror of a toric CY3 \cite{Kol:1997fv,Brandhuber:1997ua}. The periods \eqref{eq:specgeom} are then solutions of a  generalised hypergeometric system of coupled PDEs given by the Gelfand--Kapranov--Zelevinsky (GKZ) system for the corresponding toric variety. These solutions can be effectively computed asymptotically around the large complex structure/weak gauge coupling point using Frobenius' method: for $N=2$ and fundamental matter, this method was brought to fruit in \cite{Eguchi:2000fv} (see also \cite{Closset:2021lhd}). The toric condition on the engineering geometry however confines this approach to unitary gauge groups, and the case $\cG\neq \mathrm{SU}(N)$  requires fundamentally different ideas to be treated.\medskip

In the present paper we address this problem by providing
a systematic construction of Picard--Fuchs operators annihilating the periods \eqref{eq:specgeom} for a general simple gauge group $\cG$. 
We will offer two constructions of such global D-modules over the Coulomb branch, and use them to test proposals for 5d SW curves arising from geometric engineering, brane constructions, and/or the theory of relativistic integrable systems. 

\subsubsection*{Picard--Fuchs ideals from Frobenius manifolds}

Suppose first that $\cG$ is of type ADE. We propose that SW periods are, in the terminology of \cite{MR2070050}, the {\it odd} periods of the canonical Frobenius manifold structure on the orbits of the Dubrovin--Zhang extension of the affine Weyl group of type $\cG$ in the reflection representation \cite{MR1606165}, which was recently explicitly constructed in \cite{Brini:2017gfi,Brini:2021pix}. This is a natural generalisation of an idea of Dubrovin \cite{MR2070050} (see also \cite{Eguchi:1996nh,Ito:1997ur} for previous work on this), where the polynomial Frobenius manifold of type ADE played a similar role in the reconstruction of the SW periods for  four-dimensional $\cN=2$ super Yang--Mills with simply-laced gauge symmetry.
In particular we propose that the Picard--Fuchs ideal annihilating the physical periods of the SW curve can be read off from a suitable specialisation of the quantum differential equations for the associated Dubrovin--Zhang Frobenius manifold. Schematically, and in close analogy to \cite{Eguchi:1996nh,Ito:1997bd,Ito:1997ur,Ito:1998zr}, we propose that there exists a distinguished holomorphic chart $\{t_i(u)\}_{0\leq i \leq r}$ on\footnote{The second factor accounts for the complexified radius of the fifth-dimensional circle, or rather its product with the holomorphic energy scale in the gauge theory.} 
$\cM_\mathsf{C}\times \bbP^1$
such that the periods \eqref{eq:specgeom} satisfy a holonomic system of Fuchsian PDEs in the form
\bea
\left(\de_{t_0}+\sum_i \mathfrak{q}_i t_i \de_{t_i}\right)^2 \Pi &=& 4 h_{\mathfrak{g}}^2 \de_{t_r}^2 \Pi, \nn \\
\de^2_{t_i t_j} \Pi &=& \mathsf{C}_{ij}^k \de^2_{t_k t_r} \Pi 
\label{eq:PFintro}
\eea
In \eqref{eq:PFintro}, $\{\mathfrak{q}_i\}_{i=1}^r$ are  the $\alpha$-basis coefficients of the highest element of the root system of type $\cG$, $h_{\mathfrak{g}}$ is the Coxeter number, the coordinates $t_i$ are a homogeneous choice of flat coordinates for the associated affine Weyl Frobenius manifold \cite{MR1606165,Brini:2021pix}, and $\mathsf{C}_{ij}^k$ are the structure constants of the Frobenius product in those coordinates. From the engineering point of view, and as an application to local mirror symmetry/(orbifold) Gromov--Witten theory, the Picard--Fuchs ideals \eqref{eq:PFintro} specialise to the GKZ system for the associated $Y^{N,0}$ toric singularities considered in \cite{Eguchi:2000fv,Brini:2008rh} for $\cG=\mathrm{SU}(N)$. When $\cG\neq\mathrm{SU}(N)$, they further generalise them to the non-toric singularities given by orbifolds of the singular conifold by a finite group action $\Gamma \subset \mathrm{SL}(2,\bbC)$ which is McKay-dual to $\cG$.

\subsubsection*{Picard--Fuchs ideals from Jacobi rings}

When $\cG$ is non-simply-laced, a naive application of the above approach fails. The reasons are already well-known from the study of the 4d setup, where the auxiliary Frobenius manifold would arise from 
the spectral curves of the non-relativistic periodic Toda lattice. It was found shortly after the work of Seiberg--Witten \cite{Martinec:1995by} that the integrable system relevant for  $\cN=2$ super Yang--Mills theory with gauge group $\cG$ is the Toda lattice associated to the {\it twisted} Kac--Moody algebra $(\cG^{(1)})^\vee$, whereas the construction in \cite{MR1606165,Brini:2021pix} of Frobenius manifolds on orbits of extended affine Weyl groups pertains to its untwisted counterpart. Unfortunately there turns out to be no strict analogue of the Frobenius manifolds of \cite{MR1606165} in the twisted Kac--Moody world, as the would-be Frobenius metric constructed from the spectral curve becomes either degenerate, or curved, in that setting. \\
Motivated by work on associativity equations for 4d-prepotentials in \cite{Bonelli:1996qh,Marshakov:1996ae,Marshakov:1997ny}, we propose an alternative method to construct Picard--Fuchs systems in the form \eqref{eq:PFintro} for the pure five-dimensional gauge theory on $\bbR^{4}\times S^1$ in a completely algebraic fashion. Our procedure takes  as its input datum 
the polynomial $\mathsf{P}_{\cG}$ defining the five-dimensional spectral curve:
the tensor $\mathsf{C}$ is identified in this context with the structure constants of a canonical subring 
of the algebra of regular functions on the zero-dimensional scheme corresponding to the ramification locus of the curve.
Its (non-trivial) existence and explicit construction is reduced to a problem in commutative algebra, which we solve 
for all $\cG$ and for all different realisations of Seiberg--Witten geometries when more than one is available (such as for non-simply-laced classical groups). By the mirror theorem of \cite{Brini:2021pix}, for simply-laced  $\cG$ this restricts to the structure constants of the affine-Weyl Frobenius algebras mentioned previously.

%
%
Owing to the absence of an underlying Frobenius metric for non-simply-laced $\cG$,
there is now no natural notion of ``flat'' coordinates $t_i(u)$, and it is a priori unclear how to fix a privileged chart for writing down something like \eqref{eq:PFintro}. We claim that the constraints of rigid special K\"ahler geometry, and in particular the existence of a prepotential, are in fact sufficient to uniquely pin down an analogous coordinate frame $t_i(u)$ for general $\cG$, without reference to an underlying flat metric. 

\subsubsection*{Matching with the gauge theory, and an integrable systems surprise}

Using Frobenius' method, the above constructions provide an efficient way of computing the prepotential given the SW/B-model mirror curve, and therefore the putative five-dimensional prepotentials for the corresponding pure gauge theory, around any boundary point in moduli space, and in particular at infinity in the Coulomb branch. \\
We put this to the test in a wide array of cases. When the SW geometry arises from geometric engineering  \cite{Katz:1996fh,Borot:2015fxa} or brane constructions in M-theory \cite{Kol:1997fv,Hayashi:2017btw,Li:2021rqr}, we show that our constructions recover the microscopic gauge theory results with flying colours. We match the perturbative and instanton parts of the prepotential with the expressions arising from the K-theoretic version of the Nakajima--Yoshioka blow-up equations \cite{Nakajima:2005fg}, as well as with their extrapolation to general gauge groups \cite{Keller:2012da}. We also apply our proposal to the SW curves for classical non-simply-laced groups obtained from string theory  constructions with orientifold planes  \cite{Brandhuber:1997ua,Hayashi:2017btw,Li:2021rqr}. As a byproduct of our construction we
show, in type $B_n$, that the curves proposed by \cite{Brandhuber:1997ua} from an M-theory uplift of the Hanany--Witten construction with  orientifold planes correctly reproduce the instanton prepotential for $\mathrm{SO}(2n+1)$ gauge groups; and in type $C_n$, we confirm a tension between the results of \cite{Brandhuber:1997ua} for $n=1$ and more recent proposals
of $\mathrm{Sp}(1)$ curves with discrete $\theta$-angle at $\theta=\pi$ in \cite{Hayashi:2017btw,Li:2021rqr}. Although the two curves are closely related, and their periods are shown to be annihilated by the differential ideal generated by the second line of \eqref{eq:PFintro} for the same choice of structure constants $\mathsf{C}_{ij}^k$, in this rank-1 case determining the full expression for the periods requires fixing a finite-dimensional ambiguity akin to imposing a quasi-homogeneity condition as in the first line of \eqref{eq:PFintro}: our calculations show that this is done differently for the curves from \cite{Brandhuber:1997ua} on one hand and of \cite{Hayashi:2017btw,Li:2021rqr} on the other, ruling out the former and confirming the correctness of the latter with a direct prepotential calculation. We also match our B-model construction applied to the $\theta=0$ version of the brane web  geometries with an O5-plane of \cite{Hayashi:2017btw,Li:2021rqr} against the corresponding instanton calculation of the prepotential -- again finding perfect agreement.
 \medskip

We furthermore apply our construction to the spectral curves of the periodic relativistic Toda chain on the Langlands dual group of the affine Poisson--Lie group $\cG^{(1)}$ \cite{MR1993935}, which are natural relativistic deformations of the well-known four-dimensional SW curves for the $\cN=2$ theory with gauge group $\cG$ \cite{Martinec:1995by}, and which have already implicitly appeared by the usual procedure of Dynkin folding in the context of geometric engineering on local CY singularities \cite{Borot:2015fxa}. For non-simply-laced cases these are formally different from the curves obtained from string engineering -- and for exceptional $\cG$, 
they are to our knowledge the only candidates available so far for a B-model/SW description of the low energy theory. 

We show that our construction is (non-trivially) well-defined in this context as well: there is an $(r+1)$-dimensional vector subspace of the ring of regular functions of the branch locus of the SW curve, non-trivially closing under multiplication to a subring with structure constants $\mathsf{C}_{ij}^k$, $i,j,k=0,\dots, r$, and once again a distinguished chart in the sense of \eqref{eq:PFintro} is shown to exist. We then proceed to solving the resulting Picard--Fuchs system at large complex structure: for non-simply-laced groups the comparison with the gauge theory result surprisingly {\it fails} in this case, away from the divisor in the Coulomb branch corresponding to the non-relativistic/4d limit, already for the perturbative prepotential. 
We corroborate our findings with an explicit residue calculation of the triple derivatives of the 1-loop prepotential from the SW curve, which we find in agreement with the calculation from our proposed Picard--Fuchs system. It thus remains an open problem to identify the correct  integrable system counterpart  of non-simply-laced gauge theories on $\bbR^4 \times S^1$, and, for exceptional non-simply-laced groups, to determine their SW geometry.

\subsubsection*{Organisation of the paper}
The paper is structured as follows. In \cref{sec:intro5d} we give a review of instanton counting and blow-up equations in four and five dimensions one one hand, and five-dimensional Seiberg--Witten curves from string theory engineering and integrable systems on the other. 
Then, in \cref{sec:ade}, we formulate our B-model approach 
to compute the gauge theory prepotential for simply-laced Lie algebras from the odd periods of the corresponding extended affine Frobenius manifold, and present detailed tests of our proposal in low rank. In \cref{sec:bcfg} we generalise our construction to non simply-laced examples, and provide an extended set of examples supporting its validity. A summary and prospects for future work are discussed in \cref{sec:conclusion}. We collect the notation used throughout the text in \cref{tab:notation} for the reader's convenience.

\subsubsection*{Acknowledgements}
We acknowledge discussions and correspondence with G.~Bonelli, K.~Ito, H.~Kanno, A.~Klemm,  P.~Su\l kowski, A.~Tanzini, and F.~Yagi. We are especially grateful to F.~Yagi for his very helpful comments about the content of \cref{sec:c1pi}.
This project has been supported by the Engineering and Physical Sciences Research Council under grant agreement ref.~EP/S003657/2. The work of KO is also in part supported by the TEAM programme of the Foundation for Polish Science co-financed by the European Union under the European Regional Development Fund (POIR.04.04.00-00-5C55/17-00).

\newpage

\begin{table}[!h]
    \centering
    \begin{tabular}{|c|p{10cm}|p{2.5cm}|}
    \hline 
    Symbol & Meaning & First instance \\
    \hline
    \hline
         $\cG$; $\cG^{(1)}$ & The compact form of a simple, simply-connected Lie group;  the corresponding affine Lie group. & \cref{sec:intro} \\ \hline
         $\mathfrak{g}$/$\mathfrak{g}^{(n)}$ & The simple complex Lie algebra $\mathrm{Lie}_\bbC(\cG)$; the corresponding affine version in Kac notation with twisting order $n$. & \cref{sec:intro}; \cref{sec:SWtwToda}. \\ \hline
         $\cK(\mathfrak{g})/\cK(\mathfrak{g}^{(n)})$ & The Cartan matrix of the simple/(twisted)-affine Lie algebra $\mathfrak{g})/\mathfrak{g}^{(n)}$ & \eqref{eq:pbg} \\
         \hline
          $\cA(\mathfrak{g})$; $\cA(\mathfrak{g}^{(1)})$ & A simply-laced simple (resp. affine) Lie algebra, from which $\mathfrak{g}$ (resp. $(\mathfrak{g}^{(1)})^\vee$) is obtained as a quotient by outer automorphisms.
          & \cref{sec:SWtwToda} \\ \hline
                $\Delta$/$\Delta^+$/$\Delta_\ell$/$\Delta_\ell^+$ &
                 The set of all/positive/long/long positive roots of $\mathfrak{g}$
          & \cref{sec:intro5d} \\ \hline
         $\rho_{[i_1\dots i_r]} $ & The irreducible representation of $\cG$ with highest weight $\omega=[i_1\dots i_r]$ in Dynkin notation & \cref{sec:SWtwToda}\\ \hline
         $\omega_i$ (resp. $\rho_i$) & The $i^{\rm th}$ fundamental weight (resp. representation) of $\cG$ & \cref{sec:intro5d}  \\ \hline
         $\mathfrak{q}_i$, $\mathfrak{q}^\vee_i$ & The Coxeter (resp. dual Coxeter) coefficients of $\mathfrak{g}$ & \eqref{eq:PFintro} \\ \hline
         $\cF_{\cG}$, $\cF_{\cG}^{[n]}$ & The prepotential (resp. the $n^{\rm th}$ instanton contribution to it) of the $\cN=2$ pure theory with gauge group $\cG$ on $\bbR^{1,3} \times S^1_{R_5}$.   & \eqref{eq:prep5d}--\eqref{eq:prepnp} \\ \hline
         $\mathsf{R}_{\cG,\rho; u}$, $\widetilde{\mathsf{R}}_{\cG;u}$ & The characteristic (resp. reduced) polynomial of the Lax operator for the untwisted affine relativistic Toda chain 
         in the representation $\rho$
         & \eqref{eq:laxaff2} \\ \hline
         $\mathsf{T}_{\cG,\rho; u}$, $\widetilde{\mathsf{T}}_{\cG,\rho;u}$
         & The characteristic (resp. reduced) polynomial of the Lax operator for the twisted affine relativistic Toda chain in the representation $\rho$ & \eqref{eq:laxtw} \\ \hline
         $\mathsf{Q}_{\cG;u}$, $\widetilde{\mathsf{Q}}_{\cG;u}$ & The defining (resp. reduced) polynomial of the Hanany--Witten M-theory curve for type $\cG$
         & \eqref{eq:QAn}--\eqref{eq:QCnb} \\ \hline
        $\mathsf{P}_{\cG;u}$; $\widetilde{\mathsf{P}}_{\cG;u}$ & One of $\mathsf{Q}$, $\mathsf{R}$, $\mathsf{T}$; resp. $\widetilde{\mathsf{Q}}$, $\widetilde{\mathsf{R}}$, $\widetilde{\mathsf{T}}$ above  & \cref{sec:intro}  \\ \hline
                  $(u_0, \dots, u_r)$ &  Ad-invariant classical coordinates on the Coulomb branch &  \cref{sec:intro5d} \\ 
 \hline
         $(q_0, \dots, q_r)$ & Exponentiated A-model Coulomb moduli & \cref{sec:intro5d} \\ \hline
         $(z_0, \dots, z_r)$ & B-model coordinates around the large complex structure point & \eqref{eq:utoz} \\ \hline
         $(t_0, \dots, t_r)$ & Flat coordinates of $M^{\rm trig}_\mathfrak{g}$; the distinguished coordinates of \eqref{eq:PF5prodgen} & \eqref{eq:PFintro} \\ \hline
         $\cV(\widetilde{\mathsf{P}}_{\cG;u})$ & The canonical subring of the space of regular functions on the ramification locus spanned by $\{\de_{t_i} \widetilde{\mathsf{P}}_{\cG;u}/\de_{t_r} \widetilde{\mathsf{P}}_{\cG;u}\}$ & Section~\ref{sec:Jacobi alg} \\ \hline  
         $\mathrm{GB}(\widetilde{\mathsf{P}}_{\cG;u})$ & A reduced Gr\"obner basis for  $\DD(\widetilde{\mathsf{P}}_{\cG;u})$  & Section~\ref{sec:Jacobi alg} \\ \hline 
         $\mathsf{C}_{ij}^k$
         & Structure constants of $\cV(\widetilde{\mathsf{P}}_{\cG;u})$ in the $t$-
        chart & \eqref{eq:PFintro}
        \\ \hline
         $M^{\rm trig}_\mathfrak{g}$; $F^{\rm trig}_\mathfrak{g}$
         & The extended affine Frobenius manifold of type $\mathfrak{g}$; its prepotential & \cref{sec:Frobenius} \\ \hline
    \end{tabular}
    \caption{Notation employed throughout the text.}
    \label{tab:notation}
\end{table}

\newpage

\section{Gauge theories and instanton counting in five dimensions}
\label{sec:intro5d}

The field content of the minimally supersymmetric Yang--Mills theory with gauge group $\cG$ in five dimensions is given by a gauge field $A_\mu$ with a Dirac spinor $\lambda$  and a real scalar $\phi$, both in the adjoint representation of $\cG$. Upon compactification on a circle of radius $R_5$ to $\bbR^4 \times S_{R_5}^1$,
the (classical/quantum) moduli space of the theory is parametrised by the (classical/quantum) vev of the complexified scalar $\varphi \coloneqq \phi+\ri A_5$. Fixing a set of linear generators $\{h_i\}_{i=1}^r$ for the Lie algebra of the maximal torus of $\cG$ and in a diagonal gauge for $\varphi$, we shall write these respectively as
\beq 
\bra \varphi \ket_{\rm cl} = \sum_{i=1}^r a_i^{\rm cl} h_i, \quad \bra \varphi \ket = \sum_{i=1}^r a_i h_i,
\label{eq:avar}
\eeq 
and write $q_i^{\rm cl} \coloneqq \re^{2 \pi \ri R_5 a_i^{\rm cl}}$, $q_i \coloneqq \re^{2 \pi \ri R_5 a_i}$ for the corresponding exponentiated linear coordinates on the Cartan torus. An alternative set of coordinates which arises naturally in  B-model approaches to $\cN=2$ theories is given by the classical vev of (the conjugacy class of) the complexified Wilson loop
%
$g=P\exp\int_{S^1_{R_5}} (-\ri \varphi)$. A choice of $r$-independent ${\rm Ad}$-invariant holomorphic functions on $\cG$ fixes then a chart on the classical Coulomb branch, a natural one being given by the fundamental traces %
\beq
u_i \coloneqq \bra \Tr_{\rho_i} P\exp\int_{S^1_{R_5}} \left(A_5-\ri \phi\right)\ket_{\rm cl}, \quad i=1, \dots, r,
\label{eq:uvar}
\eeq 
where $\rho_i$ is the irreducible representation having  the $i^{\rm th}$ fundamental weight $\omega_i$ of $\cG$ as its highest weight. By \eqref{eq:avar} and their definition in \eqref{eq:uvar}, the $u$-coordinates are Weyl-invariant integral Laurent polynomials in $(q_1^{\rm cl}, \dots, q_r^{\rm cl})$:
\beq
u_i \in \bbZ[q_1^{\rm cl}, \dots, q_r^{\rm cl}].
\eeq 
Treating further the compactification radius $R_5$ as a free parameter, there is an additional dimensionless modulus compared to the usual four-dimensional theory given by
\beq
u_0 := \left(\frac{1}{R_5 \Lambda_{\rm QCD}}\right)^{2 h_\mathfrak{g}},
\label{eq:u0}
\eeq 
where $\Lambda_{\rm QCD} = \Lambda_{\rm UV}\re^{-\frac{1}{4 g^2_{\rm YM}(\Lambda_{\rm UV})}}$ is the dynamical gauge theory scale, and $h_\mathfrak{g}$ is the dual Coxeter number. We will often write $q_0 \coloneqq u_0^{-1}$ for its inverse, so that $q_0$ corresponds to the perturbative limit of the gauge theory. \medskip

When $R_5=\infty$, the prepotential of the theory was shown by Intriligator--Morrison--Seiberg (IMS) to be exact at one loop \cite{Intriligator:1997pq}, and it takes the  form of a cubic polynomial in the real scalars $\varphi$, 
\beq
\cF_\cG\Big|_{R_5=\infty} = \frac{1}{2 g_{\rm YM}^2} \mathsf{g}_{ij} \varphi^i \varphi^j + \frac{k_{\rm CS}}{6} \mathsf{d}_{ijk} \varphi^i \varphi^j \varphi^k+ \frac{1}{6}\sum_{\alpha \in \Delta} \theta(\alpha(\varphi)) \alpha(\varphi)^3
\label{eq:prep5d}
\eeq 
where $\mathsf{g}_{ij}$ and $\mathsf{d}_{ijk}$ are respectively the Killing pairing and the cubic Casimir form for $\mathfrak{g}$, $k_{\rm CS} \in \bbZ$ is the 5-dimensional Chern--Simons level, and $\theta(x)$ denotes Heaviside's step function. Upon compactification on $S^1_{R_5}$, the prepotential receives finite $u_0$ perturbative corrections from an infinite tower of excited Kaluza--Klein states, as well non-perturbative instanton corrections. The former are resummed as \cite{Nekrasov:1996cz}
\beq
\cF_\cG^{[0]} = -\frac{\ri}{8\pi^3} \sum_{\alpha \in \Delta} \left[
\left(2\pi \ri R_5 \alpha(a)\right)^2 \frac{\log q_0}{2}+
\mathrm{Li}_3\left(\re^{2\pi \ri R_5 \alpha(a)}\right)\right]
\label{eq:preppert}
\eeq 
%
while the latter gives rise to an infinite sum of the form
\beq
\cF_\cG^{\rm np}= \sum_{n>0} q_0^n \cF_\cG^{[n]}(q_1, \dots, q_r),
\label{eq:prepnp}
\eeq 
with $\cF_\cG^{[n]} \in (\ri/2\pi)^3 \bbQ[[q_1, \dots, q_r]]$.

\subsection{A-model: K-theoretic instanton counting and blow-up equations}
\label{sec:blowup}

We start off by giving a quick summary of instanton partition functions in four and five  dimensions, and of the generalised blow-up equations \cite{Nakajima:2005fg,Keller:2012da} that recursively determine them.

\subsubsection{Instanton counting \`a la Nekrasov: the $\mathrm{SU}(r+1)$ case}


Let $R_5 = 0$. By $\cN=2$ supersymmetry, the path integral calculation of non-perturbative contributions to the prepotential localises on instantons -- anti-selfdual connections on a principal $\cG$-bundle on $S^4 \simeq \bbR^4 \cup \{ \mathrm{pt} \}$ with fixed second Chern character, modulo gauge equivalence.
For $\cG=\mathrm{SU}(r+1)$ \cite{Nekrasov:2002qd}, this space has an algebraic compactification to  the framed moduli space $\mathcal{M}(r,n)$ 
of torsion-free sheaves on $\bbP^2$ of rank $r+1$ and $\langle c_2(E),[\bbP^2]\rangle=n$: given a pair of integers $(r,n)$, $\mathcal{M}(r,n)$ parametrises isomorphism classes of a pair $(E,\Phi)$ such that \cite{Nakajima:2003pg}:
\begin{enumerate}
    \item $E$ is a torsion-free sheaf of rank$E=r+1$ and $\langle c_2(E),[\bbP^2]\rangle=n$ which is locally free in a neighbourhood of $\ell_{\infty}=\bbP^2\backslash \bbC^2$,
    \item $\Phi:E|_{\ell_{\infty}}\xrightarrow{\sim}\cO^{\oplus r+1}$ is an isomorphism, which implies $c_1(E)=0$
\end{enumerate}
$\cM(r,n)$ is a $2n(r+1)$-dimensional smooth quasi-projective complex variety, with the open subset $\cM^{\text{reg}}_0(r,n)$ of locally free sheaves coinciding with the moduli space of instantons on $\bbS^4$ of rank $r+1$ and $c_2=n$ \cite{Donaldson:1984tm}. It also carries a $\tilde T\coloneqq (\mathbb{C}^\star)^2\times (\mathbb{C}^\star)^{r}$ action coming from the scaling actions on $\bbP^2$ and the maximal torus of $\mathrm{SU}(r+1)$, with zero-dimensional fixed loci $\cM^{\tilde T}(r,n)$. Let $\bbQ(\epsilon_1, \epsilon_2; a_1, \dots, a_r)$ be the field of fractions of $H_{\tilde T}({\rm pt}) = H(B\tilde T, \bbQ)\simeq \bbQ[\epsilon_1, \epsilon_2; a_1, \dots, a_r]$. The instanton partition function is defined by the Atiyah--Bott formula as the equivariant volume
\begin{equation}
    \cZ_{\cG}^{\text{np,4}d}\coloneqq \sum_{n\geq0}q_0^n\int_{\cM^{\tilde T}(r,n)} \frac{1}{\re_{\tilde T}\big(N_{\cM^{\tilde T}(r,n)/\cM(r,n)}\big)} \in \bbQ(\epsilon_1, \epsilon_2, a_1, \dots, a_r).
    \label{Z4d1}
\end{equation}
The localisation formula reduces the computation of $\cZ_{\cG}^{\text{np,4}d}$ to a combinatorial question involving sums of 2D-partitions, see  \cite{Nakajima:2003pg,Nakajima:2003uh} for explicit formulas.
It was conjectured in \cite{Nekrasov:2002qd} and proved in \cite{Nakajima:2003pg,Nekrasov:2003rj} that the 4d prepotential 
computed from 
the Seiberg-Witten curve coincides with the logarithm of the instanton partition function in the non-equivariant/flat $\Omega$-background limit,
$    \cF_\cG^{\text{4d}}=\lim_{\epsilon_1,\epsilon_2\to0}{\epsilon_1\epsilon_2}\log \cZ_{\cG}^{\text{np,4d}}$.
%
A natural generalisation to five compactified dimensions was also addressed by Nekrasov \cite{Nekrasov:2002qd} and later proven by Nakajima--Yoshioka \cite{Nakajima:2005fg}, by uplifting \eqref{Z4d1} to equivariant $K$-theory: 
\begin{align}
    \cZ_{\cG}^{\text{np,5}d}=&\sum_{n\geq0}\left(q_0 \re^{-rR_5(\epsilon_1+\epsilon_2)/2}\right)^nZ^{\text{np,5}d}_{\cG,n},\\ Z^{\text{np,5}d}_{\cG,n}=&\,\sum_i (-1)^i \text{ch}\, H^i(\cM(r,n),\cO),\label{Z5d1}
\end{align}
where 
ch denotes the Hilbert series \cite[Section 4.1]{Nakajima:2003pg}. Accordingly, the nonperturbative part of the prepotential \eqref{eq:prepnp} is given by
\begin{equation}
    \cF_\cG^{\text{np}}=\lim_{\epsilon_1,\epsilon_2\to0}\epsilon_1\epsilon_2\log \cZ_{\cG}^{\text{np,5d}}.\label{FandZ5d}
\end{equation}

\subsubsection{Blow-up equations}
An efficient tool to evaluate \eqref{Z5d1} was put forward by \cite{Nakajima:2003pg}, in the form of a 
comparison formula between the instanton partition function on $\bbP^2$ and the one on successive blow-ups at points. 
The upshot of that is a recursive relation for $ \cZ_{\cG}^{\text{np,5}d}$ in terms of instanton numbers $n$. 
Let $\mathbb{F}_1$ be the first Hirzebruch surface, 
\begin{equation}
    \mathbb{F}_1=\{[z_0,z_1,z_2],[x,y]\in\bbP^2\times\bbP^1|z_1y=z_2x\}.
\end{equation}
We denote by $C$ the exceptional divisor defined by $z_0=0$, by $\cO(C)$ the line bundle associated with the divisor $C$, and by $\cO(m C)$ the $m$-th tensor power for some $m\in\bbZ_{\geq0}$, and consider the framed moduli space $\widehat \cM(r,n,k)$ of torsion-free sheaves $(E,\Phi)$ on
$\mathbb{F}_1$,
where
\begin{equation}
    \bra c_1(E),[C]\ket=-k,\;\;\;\;\bra c_2(E)-\frac{r}{2(r+1)}c_1(E)^2,[\bbF_1]\ket=n.
\end{equation}

Although $n$ is not an integer in general, $\widehat \cM(r,n,k)$ is 
nonsingular of dimension $2n(r+1)$, and it carries a $\bbC^2\times \bbC^r$-torus action as in the previous section.
The corresponding $K$-theoretic instanton partition function for $\mathbb{F}_1$ is 
\begin{equation}
    \hat Z_{\mathrm{SU}(r+1),m,k}^{\text{np},5d}=\sum_{n\geq0}\left(q_0 \re^{-rR_5(\epsilon_1+\epsilon_2)/2}\right)^n \text{ch}\, \sum_i (-1)^i H^i(\widehat \cM(r,n,k),\cO(mC)).\label{Z5d2}
\end{equation}
and is defined for each tensor power $\cO(m C)$. 
In terms of $\hat Z_{\mathrm{SU}(r+1),m,k}$, Nakajima--Yoshioka \cite{Nakajima:2005fg} establish a finite difference equation in the equivariant parameters and Coulomb moduli, as follows.
%
Let $\cG=\mathrm{SU}(r+1)$ and $\vec{a}\in \bbR^{r+1}
$ with $\sum_{i}a_i=0$ be an element of the Cartan subalgebra of $\mathfrak{g}=\mathfrak{sl}_{r+1}$, and further write $K^{\vee}$ for the special linear coroot lattice, $K^{\vee}=\{\vec{k}\in\mathbb{Z}^{r+1}|\sum_ik_i=0\}$. Then for $k=0$, $\hat Z_{\cG,m,0}^{\text{np},5d}$ satisfies
\begin{align}
   \hat Z_{\cG,m,0}^{{\rm np},5d}=&\sum_{\vec{k}\in K^{\vee}}\frac{(q_0 \re^{R_5(\epsilon_1+\epsilon_2)(m-h_{\mathfrak{g}}^{\vee}/2)})^{(\vec{k},\vec{k})/2}e^{(\vec{k},\vec{a})m}}{\prod_{\alpha\in\Delta}l_{\alpha}^{\vec{k}}(\epsilon_1,\epsilon_2,\vec{a})}\nonumber\\
    &\times Z_{\cG}^{{\rm np},5d}(\epsilon_1,\epsilon_2-\epsilon_1,\vec{a}+\epsilon_1 \vec{k};e^{R_5\epsilon_1(m-h_{\mathfrak{g}}^{\vee}/2)}q_0,R_5)\nonumber\\
    &\times Z_{\cG}^{{\rm np},5d}(\epsilon_1-\epsilon_2,\epsilon_2,\vec{a}+\epsilon_2 \vec{k};e^{R_5\epsilon_2(m-h_{\mathfrak{g}}^{\vee}/2)}q_0,R_5),\label{blowup1}
\end{align}
where
\begin{equation}
l_{\alpha}^{\vec{k}}(\epsilon_1,\epsilon_2,\vec{a})=
\begin{dcases}
    \hspace{7mm}\prod_{\substack{i,j>0\\i+j\leq-(\vec{k},\vec{a})-1}}\left(1-e^{R(i\epsilon_1+j\epsilon_2-(\vec{a},\alpha))}\right) & {\rm if}\;\; (\vec{k},\alpha)<0,\\
   \prod_{\substack{i,j>0\\i+j\leq(\vec{k},\vec{a})-2}}\left(1-e^{R(-(i+1)\epsilon_1-(j+1)\epsilon_2-(\vec{a},\alpha))}\right) & {\rm if}\;\; (\vec{k},\alpha)>1,\\
   \hspace{35mm}1 & {\rm otherwise}
\end{dcases}
\end{equation}
Furthermore, for all $m\in\{0,1,...,r+1\}$, we have that 
\begin{equation}
    \hat Z_{\cG,m,0}^{{\rm np},5d} = Z_{\cG}^{{\rm np},5d},\label{blowup2}
\end{equation}
Note that $h_{\mathfrak{sl}_{r+1}}^{\vee}=r+1$ for $\mathrm{SU}(r+1)$. 
From \eqref{blowup1}, it is possible to obtain a recursion for $Z_{\cG,n}^{\text{np,5}d}$, i.e., the expansion coefficients in terms of $q_0$ \eqref{Z5d2}:  see \cite{Nakajima:2005fg}.

\subsubsection{General gauge groups}

The discussion above is restricted to the case of $\cG=\mathrm{SU}(r+1)$, and except for the classical gauge groups where an ADHM construction is known \cite{Nekrasov_2004,Marino_2004,Fucito_2004}, it is a priori unclear how to deduce an immediate generalisation of \eqref{blowup1} for other Lie types. However it is not hard to extrapolate \eqref{blowup1} to general simple, simply-connected $\cG$ as the r.h.s. is entirely written in terms of purely root-theoretic data \cite{Keller:2012da}: indeed, replacing the type~A root system of \eqref{blowup1} with the one of an arbitrary simple Lie algebra $\mathfrak{g}$ leads to formulas for 1-instanton partition which are consistent with supersymmetric index calculations \cite{Benvenuti_2010,Gaiotto:2009jjh,Keller_2012}. 
By taking these results into consideration, the authors of \cite{Keller:2012da} derived another recursion for $Z_{\cG,n}^{\text{np,5}d}$ from \eqref{blowup1} in the following form:
\begin{equation}
    Z_{\cG,n}^{{\rm np},5d}=\frac{\re^{R_5n(\epsilon_1+\epsilon_2)}I_n^{(0)}-\left(\re^{R_5n\epsilon_1}+\re^{R_5n\epsilon_2}\right)I_n^{(1)}+I_n^{(2)}}{(1-\re^{R_5n\epsilon_1})(1-\re^{R_5n\epsilon_2})},\label{KS1}
\end{equation}
where
\begin{align}
    I^{(m)}_n=&\sum_{\substack{\vec{k}\in K^{\vee}\\i,j<n}}^{\frac12(\vec{k},\vec{k})+i+j=n}\frac{\exp\left(R_5m\left(i\epsilon_1+j\epsilon_2+(\vec{k},\vec{a})+(\vec{k},\vec{k})\frac{\epsilon_1+\epsilon_2}{2}\right)\right)}{\prod_{\alpha\in\Delta}l_{\alpha}^{\vec{k}}(\epsilon_1,\epsilon_2,\vec{a})}\nonumber\\
    &\times Z_{\cG,i}^{{\rm np},5d}(\epsilon_1,\epsilon_2-\epsilon_1,\vec{a}+\epsilon_1 \vec{k},R_5) \,Z_{\cG,j}^{{\rm np},5d}(\epsilon_1-\epsilon_2,\epsilon_2,\vec{a}+\epsilon_2 \vec{k},R_5).\label{KS2}
\end{align}
Note that $I^{(m)}_n$ is determined by $ Z_{\cG,n'}^{{\rm np},5d}$ with $n'<n$, and $ Z_{\cG,0}^{{\rm np},5d}=1$.
In particular, denoting $\Delta_{\ell}$ the set of long roots, we get
\begin{align}
Z_{\cG,1}^{{\rm np},5d}(\epsilon_1,\epsilon_2,\vec{a})=&\frac{1}{(1-\re^{-\epsilon_1})(1-\re^{-\epsilon_2})}\nonumber\\
&\sum_{\gamma\in\Delta_{\ell}}\frac{1}{(1-\re^{-\epsilon_1-\epsilon_2+\gamma\cdot a})(1-\re^{-\gamma\cdot a})\prod_{\alpha\cdot\gamma=1}(1-\re^{-\alpha\cdot a})},
\end{align}
from which we find
\begin{align}
\cF_{\cG}^{[1]}=\sum_{\gamma\in\Delta_{\ell}}\frac{1}{(1-\re^{\gamma\cdot a})(1-\re^{-\gamma\cdot a})\prod_{\alpha\cdot\gamma=1}(1-\re^{-\alpha\cdot a})},\label{1instanton}
\end{align}
where we used \eqref{FandZ5d}. 

\subsection{B-model: Seiberg--Witten geometries, old and new}

We move on to review the B-model counterpart of instanton counting, given by special geometry on families of spectral curves fibred over the Coulomb branch. We discuss in turn the two main sources of these for five-dimensional gauge theories, namely M-theory engineering and the spectral curves from relativistic integrable models. The two approaches lead to a priori inequivalent B-model geometries for non simply-laced groups, as we now review.

\label{sec:SWcurves}

\subsubsection{Spectral curves from M-theory engineering}
\label{sec:SWMth}

String theory embeddings are an extremely helpful tool to analyse the infrared behaviour of supersymmetric quantum field theories. For the case of eight supercharges, it is well known that one can realise SUSY gauge theories with gauge symmetry given by products of classical Lie groups and bifundamental matter as the low energy theory on systems of D-branes suspended between NS5-branes \cite{Hanany:1996ie,Witten:1997sc,Kol:1997fv}. In the strong type II string coupling limit, the resulting M-theory picture is given by a single smooth fivebrane setup, containing the Seiberg--Witten curve as a factor \cite{Witten:1997sc,Brandhuber:1997ua}. 

For 5d theories with a single $\cG=\mathrm{SU}(N)$ factor and no hypermultiplets, one considers
a IIB configuration on $\bbR^{1,3} \times S^1_{R_5} \times \bbR \times \bbR^4$ with $N$ light D5-branes located at $x^i=0$, $i=5,7,8,9,10$, sweeping a full $\bbR^{1,3} \times S^1_{R_5}$ in the $x^0, \dots x^4$ directions and suspended between two non-dynamical solitonic fivebranes classically situated at finite distance in the $x^6$ coordinate; the latter are extended in the $x^0, \dots, x^5$ coordinates. The theory on the D5-branes is macroscopically five-dimensional, owing to the the finite stretch in the $x^6$ direction. This brane setup is T-dual to the constructions considered for theories with eight supercharges in 3d and 4d respectively by \cite{Hanany:1996ie} and \cite{Witten:1997sc}, and it admits an M-theory description on $\bbR^{1,3} \times S^1_{R_5} \times \bbR \times \bbR^4 \times S^1_{R_{11}}$ in terms of an M5 brane wrapping a supersymmetric cycle $\bbR^{1,3} \times \cC$, for a non-compact Riemann surface $\cC$. In terms of the $\bbC^\star$-coordinates $\lambda=\re^{-\ri (x^4 + \ri x^5)/R_5}$  and $\mu= \re^{-(x^6 + \ri x^{10})/R_{11}}$, $\cC$ satisfies an algebraic equation of the form 
\beq
\mathsf{Q}_{\mathrm{SU}(N); u}(\mu,\lambda) = \lambda^2 + Q_1(\mu) \lambda + Q_2(\mu) = 0
\eeq 
where roots of $Q_1(\mu)$ (resp. $Q_2(\mu)$) correspond, in the type IIB limit, to the location of D5-branes between NS5s (resp. at $x^6=\infty$, corresponding to fundamental hypermultiplets in the gauge theory on the D5 worldvolume). For the case of no hypermultiplets and Chern--Simons level $k_{\rm CS}\in \{0,\dots, N\}$, one finds \cite{Brandhuber:1997ua}
\beq 
\label{eq:QAn}
\mathsf{Q}_{\mathrm{SU}(N); u}(\mu,\lambda) = \lambda^2+
\lambda\left[1+\sum_{k=1}^{N-1} u_k \mu^{k} + \mu^{N}\right] + q_0 \mu^{N-k_{\rm CS}}.
\eeq 

%
Other classical gauge groups can be analysed in the same manner through the addition of orientifold O5-planes in the type IIB picture. This again can be subsumed into an M-theory description in terms of a fivebrane with worldvolume containing the affine part $\{(\mu,\lambda) \in \bbC^\star \times \bbC^\star | \mathsf{Q}_{\cG;u}(\mu,\lambda)=0\}$ of a hyperelliptic Riemann surface. In the special orthogonal case, this yields \cite{Brandhuber:1997ua}
\bea
\label{eq:QBn}
\mathsf{Q}_{\mathrm{Spin}(2N+1); u}(\mu,\lambda) &=& \mu^{2N-3}\left(\mu^2-1\right) \left(\mu^2+1\right)^2 (\lambda^2+q_0) + \lambda\left[1+\sum_{k=1}^{2N-1} u_k \mu^{2k} + \mu^{4N}\right]
,  \\
\label{eq:QDn}
\mathsf{Q}_{\mathrm{Spin}(2N); u}(\mu, \lambda) &=& \mu^{N-2}  (\mu^2-1)^2\left(\lambda^2+q_0\right) + \lambda \left[1+\sum_{k=1}^{2N-1} u_k \mu^k+\mu^{2N}\right].
\eea 
where 
$u_{k}=u_{2N-k}$. 
For symplectic gauge groups, since $\pi_4(\mathrm{Sp}(N))=\bbZ/2$, the gauge theory path integral depends on an additional discrete ambiguity, which we can regard as a 5d analogue of a theta angle $\re^{\ri \theta}$ where $\theta \in \{0,\pi\}$. The difference between the two choices can be reabsorbed in a rescaling of the masses of the fundamental hypermultiplets when present, but it is physical for the pure gauge theory, leading to inequivalent prepotentials. Accordingly, two sets of curves $\{\mathsf{Q}_{\mathrm{Sp}(N); u,\theta}=0\}$ were put forward in 
\cite{Hayashi:2017btw,Li:2021rqr} to account for this,
\bea 
\label{eq:QCny0}
\mathsf{Q}_{\mathrm{Sp}(N); u,0}(\mu,\lambda) &=& \mu^{N+2} \left(\lambda^2+ q_0\right)+ \lambda \Bigg[ \sum_{k=2}^{N+2} u_k \left(\mu^{N+2+k}+\mu^{N+2-k}\right) 
\nn \\
&+& \left(q_0-1-\sum_{k=1}^{\frac{N-1}{2}}u_{2k+1}\right) \left(\mu^{N+3}+\mu^{N+1}\right) -2 \sum_{k=1}^{\frac{N+1}{2}} u_{2k}\Bigg],
 \\ 
\label{eq:QCnypi}
\mathsf{Q}_{\mathrm{Sp}(N); u,\pi}(\mu,\lambda) &=& 
\mu^{N+2} \left(\lambda^2+ q_0\right)+ \lambda \Bigg[ \sum_{k=2}^{N+2} u_k \left(\mu^{N+2+k}+\mu^{N+2-k}\right) 
\nn \\
&-& \left(1+\sum_{k=1}^{\frac{N-1}{2}}u_{2k+1}\right) \left(\mu^{N+3}+\mu^{N+1}\right) -2 \left(q_0+ \sum_{k=1}^{\frac{N+1}{2}} u_{2k}\right)\Bigg].
\eea 
with $u_{N+2}=1$. A pre-existing set of curves for the case with $N_f\leq N+2$ fundamental flavours was also proposed in \cite{Brandhuber:1997ua}, although the discussion of the $\theta$-angle dependence was not carried out. Naively setting $N_f=0$ leads to an alternative M-theory curve whose defining polynomial we denote by $\mathsf{Q}_{\mathrm{Sp}(N); u}^\flat$, where
\beq 
\mathsf{Q}_{\mathrm{Sp}(N); u}^\flat (\mu,\lambda) = \mu^{N+2}\left(\lambda^2+ q_0\right)+
\lambda(\mu^2-1)^2\left[1+\sum_{k=1}^{2N-1} u_k \mu^{k} + \mu^{2N}\right]
\label{eq:QCnb}
\eeq 
and $u_{k}=u_{2N-k}$. We shall in fact see in \cref{sec:c1pi} in the example $N=1$ that 
the prepotential computed from \eqref{eq:QCnb} 
disagrees with the $\mathrm{Sp}(1)$ gauge theory prepotential 
for either choice of discrete $\theta$-angle.

\subsubsection{Spectral curves from integrable systems}
\label{sec:SWtwToda}

Another, and historically the first \cite{Nekrasov:1996cz}, main source of spectral curves for gauge theories in five compactified dimensions comes from the theory of relativistic integrable systems; and in particular, for the pure gauge theory, from the study of the affine relativistic Toda chain \cite{Nekrasov:1996cz,MR979202, MR1090424}. We refer to \cite{Nekrasov:1996cz, Borot:2015fxa, Brini:2017gfi, Fock:2014ifa, Williams:2012fz, MR1993935} for background and further details, condensing here the information relevant for the constructions of \cref{sec:bcfg}.

The phase space of the relativistic Toda chain associated to the (untwisted) affine Lie group $\cG^{(1)}$ is the  $(2r+2)$-dimensional Poisson algebraic torus $ \left(
(\bbC_x^\star)^{r+1} \times (\bbC_y^\star)^{r+1}, \{,\}_\cG\right)$ with log-constant Poisson bracket
\beq
\{x_i, y_j\}_\cG \coloneqq  \cK(\mathfrak{g}^{(1)})_{ij} x_i y_j.
\label{eq:pbg}
\eeq
Since $\cK(\mathfrak{g}^{(1)})$ has a 1-dimensional kernel, \eqref{eq:pbg} has a single Casimir function, given by 
\beq
u_0 \coloneqq \prod_{i=0}^r x_i^{-2\mathfrak{q}_i},
\label{eq:aleph}
\eeq
with $\mathfrak{q}_i$ the Coxeter exponents of $\mathfrak{g}^{(1)}$. Dynamical commuting flows are specified by the spectral-parameter-dependent Lax operator 
\cite{Fock:2014ifa,Kruglinskaya:2014pza, MR1993935},
\bea
L_\cG(x,y;\lambda) &=& 
E_0(\lambda/y_0) E_{\bar 0}(\lambda) \prod_{i=1}^r H_i(x_i) E_i(1)
H_i(y_i) E_{-i}(1)
\label{eq:laxaff2}
\eea
where $E_i(x)=\exp(x e_i)$, $H_i(x)=\exp(x h_i)$, and $\{e_{\pm i},h_i\}_{i=0,\dots, r}$ is a Cartan--Weyl basis for $\mathfrak{g}^{(1)}$. \medskip 

Let $\rho_i$, $i=1,\dots, r$ be the $i^{\rm th}$ fundamental representation of $\cG$. Then, proceeding as in \cite[Lem.~2.2]{Brini:2017gfi}, one finds that the $\lambda$-dependence of the fundamental traces of the Lax operator \eqref{eq:laxaff2} are given by
\beq
\Tr_{\rho_i} L_\cG(x,y;\lambda) = u_i(x,y) + 
\left\{
\bary{lcr}
\left(\lambda \delta_{iN}+\frac{\delta_{i,N+1}}{u_0\lambda}\right), & \qquad & \cG= \mathrm{SU}(2N+1) \\
\left(\lambda + \frac{1}{u_0\lambda}\right) \delta_{i\bar k}, & \qquad & \mathrm{else,}
\eary
\right. 
\label{eq:lambdadep}
\eeq
where $\{u_i\}_{i=1}^r$ is a complete set of commuting first integrals w.r.t. the Poisson structure \eqref{eq:pbg}, and $\bar k$ labels the fundamental weight $\omega_{\bar k}$ corresponding to the largest irreducible Weyl orbit. Fixing a non-trivial representation $\rho'\in \mathrm{Rep}(\cG)$, these integrals  determine, and can be retrieved from, the characteristic Laurent polynomial
\bea
\mathsf{R}_{\cG,\rho';u}(\mu,\lambda) & \coloneqq & \det_{\rho'} \left[ L_\cG(x,y;\lambda)-\mu\right] \nn \\ &=&\sum_{i=0}^{\dim \rho'} (-\mu)^i \Tr_{\wedge^{\dim\rho'-i}\rho}  L_\cG(x,y;\lambda) \nn \\
&\in & \bbZ[\lambda^\pm, \mu; u_0, \dots, u_r]
\label{eq:untwcharpol}
\eea 
where we have used \eqref{eq:lambdadep} and the fact that $\mathrm{Rep}(\cG)$ is an integral polynomial ring in the fundamental characters. Note that, for $\cG\neq \mathrm{SU}_{2N+1}$, \eqref{eq:lambdadep} further ensures that there exists 
$\widetilde{\mathsf{R}_{\cG,\rho';u}}(\mu, \nu)\in \bbZ[\mu, \nu; u_0, \dots, u_r]$ such that
\beq
\widetilde{\mathsf{R}}_{\cG,\rho';u}(\mu,\nu)\bigg|_{\nu=\lambda+1/(u_0\lambda)}=\mathsf{R}_{\cG,\rho';u}(\mu,\lambda).
\label{eq:redpolR}
\eeq 

As $(u_1, \dots, u_r)$, from \eqref{eq:lambdadep}, are Weyl-invariant coordinates on the maximal torus of $\cG$,  the vanishing locus of $\mathsf{R}_{\cG;u}$ gives (after normalisation of the fibres) a family of spectral curves over the Coulomb branch. For classical simply-laced groups\footnote{For the special unitary gauge group $\cG=\mathrm{SU}(N)$, the equivalence is realised at Chern--Simons level 
$k_{\rm CS}=0$. 
The effect of a shift in the Chern--Simons level on the integrable chain is discussed in \cite{Eager:2011dp,Marshakov:2019vnz}.} $\cG=A_r$ or $D_r$ one recovers \cite{Borot:2015fxa,Kruglinskaya:2014pza}, in particular, the M-theory curves \eqref{eq:QAn} and \eqref{eq:QDn}:
\begin{align}
    \mathsf{R}_{\mathrm{SU}(r+1),\rho_1;u}&=\mathsf{Q}_{\mathrm{SU}(r+1);u}\Bigr|_{k_{{\rm CS}}=0},\;\;\;\;
\mathsf{R}_{\mathrm{Spin}(2r),\rho_1;u}=\mathsf{Q}_{\mathrm{Spin}(2r);u}.
\end{align} 
The spectral polynomials $\mathsf{R}_{E_n,\rho_{n-1}}$ for $\cG=E_n$ were computed\footnote{Although the curves depend on the choice of the representation $\rho'$ in \eqref{eq:laxaff2}, the prepotentials computed for different choices of $\rho'$ only differ by an overall normalisation. The representation-independence of the special geometry prepotential is non-trivial, and follows from an isomorphism of the flows of the underlying integrable system when projected to a canonical Prym--Tyurin subvariety of the Jacobian: see \cite{Brini:2017gfi,MR1668594} for an extended discussion.} in \cite{Borot:2015fxa,Brini:2017gfi}. \medskip

\begin{table}[t]
    \centering
    \begin{tabular}{|c|c|c|}
    \hline
    Type & Folding & Dynkin diagram\\
    \hline
    \hline
    $B_l^\vee=A^{(2)}_{2l-1}$ & 
    \dynkin[edge length=.75cm,
involutions={
1{8};
27;
36;
45;}
]A[1]{****.****}
    & \dynkin[%
edge length=.75cm,
]A[2]{odd}
    \\
    \hline 
    $C_l^\vee=D^{(2)}_{l+1}$ & \dynkin[edge length=.75cm,
involutions={
[relative]0{9};
1{10};
28;
37;}
]D[1]{*****.*****}
    & \dynkin[%
edge length=1cm,
]D[2]{}  \\
    \hline 
    $F_4^\vee=E^{(2)}_{6}$ & \dynkin[%
edge length=.75cm,
involutions={07;16;35}]E[1]7 & \dynkin[%
edge length=1cm,
]E[2]6\\
    \hline 
    $G_2^\vee=D^{(3)}_{4}$ & \dynkin[%
edge length=.75cm,
involutions={01;16;60;35;52;23}]E[1]6 & \dynkin[%
edge length=1cm,
]D[3]4\\
\hline 
    \end{tabular}
    \caption{Dynkin diagrams and foldings for twisted affine Lie algebras.}
    \label{tab:twistedla}
\end{table}
When $\cG$ is of BCFG type, however, it is known that the prepotentials of the spectral curves $\{\mathsf{R}_{\cG,\rho';u}=0\}$ are not expected to reproduce the low energy effective action of the gauge theory. The reason for this was highlighted in \cite{Martinec:1995by}: in the $R_5 \to 0$ limit, corresponding to the limit of infinite speed of light of the relativistic Toda chain, the relevant dynamical system for $\cN=2$ super Yang--Mills is the affine non-relativistic Toda system associated to the {\it twisted} Kac--Moody algebras $(\mathfrak{g}^{(1)})^\vee$ \cite{MR1104219,Olive:1982ye}, rather than to $\mathfrak{g}^{(1)}$ itself.
These are given by quotients of a ``parent'' untwisted Kac--Moody algebra $\cA(\mathfrak{g}^{(1)})$
by their outer automorphism group $\mathrm{Out}(\cA(\mathfrak{g}^{(1)}))$, associated to the folding of the Dynkin diagram of $\mathfrak{g}^{(1)}$: see \cref{tab:twistedla}. 

%
%
Accordingly, it is natural to cast the twisting constructions relevant for the affine Lie algebras/ non-relativistic Toda chains/$\cN=2$ $d=4$ theories at the level of the corresponding affine Lie groups/relativistic Toda chains/$\cN=2$ KK theories, by considering the {\it twisted} Lax operator%
\beq
L_\cG^{\vee}(x,y;\lambda) \coloneqq 
E^\vee_0(\lambda/y_0) E^\vee_{\bar 0}(\lambda)\prod_{i=1}^r H^\vee_i(x_i) E^\vee_i(1)
H^\vee_i(y_i) E^\vee_{-i}(1) 
\label{eq:laxtw}
\eeq
obtained from \eqref{eq:laxaff2} by replacing all Chevalley operators with their twisted counterparts. Let $\cA(\cG) = \exp\cA(\mathfrak{g}))$: fixing $\mathbf{1} \neq \rho' \in \mathrm{Rep}(\cA(\cG))$, the associated characteristic polynomial is defined as in \eqref{eq:untwcharpol}
\bea
\mathsf{T}_{\cG,\rho';u}(\mu,\lambda) & \coloneqq & \det_{\rho'} \left[ L^\vee_\cG(x,y;\lambda)-\mu\right] \nn \\ &=&\sum_{i=0}^{\dim \rho'} (-\mu)^i \Tr_{\wedge^{\dim\rho'-i}\rho}  L^\vee_\cG(x,y;\lambda)
\label{eq:polT}
\eea 
To determine the spectral dependence of $\mathsf{T}_{\cG,\rho';u}$ on the $\lambda$ parameter, let $\mathrm{Eff}(\cA(\mathfrak{g}))$ be the set of nodes in the Dynkin diagram of the ``parent'' simple Lie algebra $\cA(\mathfrak{g})$ acted upon effectively by diagram automorphisms. This set of vertices splits in orbits $\mathrm{Eff}(\cA(\mathfrak{g}))=\sqcup_j \mathsf{o}^{(2)}_j \sqcup_k \mathsf{o}^{(3)}_k$, where $\mathsf{o}^{(n)}_i$ is a cardinality-$n$ set of Dynkin nodes upon which the outer automorphisms acts as the permutation group $S_n$ in $n$-elements. Let $\sigma : \mathsf{o}^{(n)}_j \to \bbZ/n\bbZ$ be a bijection onto the set of $n^{\rm th}$ roots of unity $\{\omega_n^l\}_{l=0}^{n-1}$, such that if $p$ is the full cyclic permutation of order $n$ in $S_n$, then $\sigma(p(\mathsf{v}))=\re^{2\pi\ri/n}\sigma(\mathsf{v})$. We claim that, up to a rescaling and a shift of $\log\lambda$, the analogue of \eqref{eq:lambdadep} in the twisted case is
\beq 
\Tr_{\rho_i} L^\vee_\cG(x,y;\lambda) = u_{\pi(i)}(x,y) + 
\left\{
\bary{lr}
\sigma(\mathsf{v}_i) \lambda + \frac{1}{u_0\lambda \sigma(\mathsf{v}_i)}, & \quad \mathsf{v}_i \in \mathrm{Eff}(\cA(\mathfrak{g})), \\
0 & \mathrm{else},
\eary 
\right. 
\label{eq:lambdatw}
\eeq 
where now the Casimir function is expressed in terms of the {\it dual} Coxeter exponents
\beq
u_0 \coloneqq \prod_{i=0}^r x_i^{-2\mathfrak{q}^\vee_i},
\label{eq:alephtw}
\eeq
%
and $\pi : \{\mathsf{v}_i\} \to \{1,\dots, r\}$ is a choice of label on the orbits, i.e. it is defined up to permutation by $\pi(\mathsf{v}_i)=\pi(\mathsf{v}_j)$ iff $\{\mathsf{v}_i,\mathsf{v}_j\} \subset \mathsf{o}^{(n)}_l$ for some $n$ and $l$. Fron \eqref{eq:lambdatw}, we can define as we did for \eqref{eq:redpolR} the reduced characteristic polynomial
\beq
\widetilde{\mathsf{T}}_{\cG,\rho';u}(\mu,\nu)\bigg|_{\nu=\lambda+1/(u_0\lambda)}=\mathsf{T}_{\cG,\rho';u}(\mu,\lambda).
\label{eq:redpolT}
\eeq 
%

\begin{example}
Let us consider the example of $\cG=B_2$, so that $\big(B_2^{(1)}\big)^\vee = \big(C_2^{(1)}\big)^\vee = D_3^{(2)}= A_3^{(2)}$.
In this case the folding of the Dynkin diagram of the simple Lie algebra $\cA(\mathfrak{b}_2)=\mathfrak{d}_3\simeq \mathfrak{a}_3$ identifies the vertices $v_1 \leftrightarrow v_3$, corresponding to the highest weights $[010]$ and $[001$] of the two Weyl spinor representations of $D_3$ (equivalently, the two complex-conjugate vector representations of $A_3$) leaving fixed $v_2$, which labels the 6-dimensional defining representation $\rho_{[100]}$ of $\mathrm{SO}(6)$ (resp. the adjoint representation of $\mathrm{SU}(4)$), so that $\mathrm{Eff}(\mathfrak{b}_2)=\mathsf{o}^{(2)}_1=\{v_1,v_3\}$: see \cref{fig:dynkB2}.

\medskip
\begin{figure}[h]
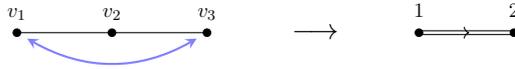
 
\centering 
\dynkin[%
edge length=1.25cm,
labels*={v_1,v_2,v_3},
involution/.style={blue!50,stealth-stealth,thick},
involutions={13}]A3
$\qquad \longrightarrow \qquad$
\dynkin[%
edge length=1.25cm,
labels*={1,2},
]B2
\caption{The Dynkin diagram of $B_2\simeq C_2$ via $A_3\simeq D_3$-folding.}
\label{fig:dynkB2}
\end{figure} 
For this example, let $\rho'=\rho_{[010]}=\mathbf{6}_\mathsf{v}$ be the 6-dimensional fundamentl representation of $A_3$ (i.e. the vector representation of $D_3=\mathrm{Spin}(6)$) and let $\varepsilon_{kl}\in \mathrm{Mat}(6,\bbC)$ with $(\varepsilon_{kl})_{ij}=\delta_{ik}\delta_{lj}$. Then the twisted Cartan--Weyl generators are given by \cite[Sec.~21.6]{MR1993935}
\beq 
e^\vee_1=\varepsilon_{21}-\varepsilon_{65}, \quad e^\vee_2=\varepsilon_{32}-\varepsilon_{53}, \quad  e^\vee_0=\varepsilon_{14}-\varepsilon_{46},
\eeq 
with $e_{\bar i}^\vee=(e_i^\vee)^T$ and $h_i = [e_i, e_{\bar i}]$, $i=0,1,2$. Then, from \eqref{eq:laxtw}, we find up to an affine transformation $\log \lambda \to a \log \lambda+b$ that
\beq 
\de_{\lambda} \Tr_{\wedge^i \rho'} L^\vee_{B_2}(x,y;\lambda) = 
 \{0,1,-2\}_i \left(2\lambda -\frac{2}{u_0 \lambda^3}\right),
\eeq 
and $\Tr_{\wedge^i \rho'} L^\vee_{B_2}=\Tr_{\wedge^{6-i} \rho'} L^\vee_{B_2}$ by the reality of $\rho'$. Using the character relations in the exterior algebra 
\beq 
\wedge^2\rho' \oplus \rho_{[000]}= \rho_{[100]} \otimes \rho_{[001]}, \quad \wedge^3\rho' = \rho_{[200]} \oplus \rho_{[002]}, 
\label{eq:wedgeB2}
\eeq 
plus the fact that $\rho_{[200]}\oplus \rho_{[010]}=\rho_{[100]}\otimes \rho_{[100]}$, $\rho_{[002]}\oplus\rho_{[010]}=\rho_{[001]}\otimes \rho_{[001]}$, and restricting to the invariant locus under the folding action, we retrieve the spectral parameter dependence claimed in \eqref{eq:lambdatw}:
\bea
\Tr_{\rho_1} L^\vee_{B_2}(x,y;\lambda) &=& u_1(x,y) + \left(\lambda+\frac{1}{u_0\lambda}\right), \nn \\
\Tr_{\rho_3} L^\vee_{B_2}(x,y;\lambda) &=& u_1(x,y) - \left(\lambda+\frac{1}{u_0\lambda}\right), \nn \\
\Tr_{\rho_2} L^\vee_{B_2}(x,y;\lambda) &=& u_2(x,y).
\label{eq:lambdaB2}
\eea 
From \eqref{eq:wedgeB2}-\eqref{eq:lambdaB2},  the reduced spectral polynomial in the 6-dimensional vector representation $\rho'=\mathbf{6}_\mathsf{v} \in \mathrm{Rep}(D_3)$ reads
\bea 
\widetilde{\mathsf{T}}_{B_2,\mathbf{6}_\mathsf{v};u}(\mu,\nu) &=&
1-u_2 \mu+\left(u_1^2-1\right) \mu^2-2 \left(u_1^2-u_2\right) \mu^3 \nn \\ 
&+& \left(u_1^2-1\right) \mu^4-u_2 \mu^5+\mu^6-\nu^2 (\mu+1)^2 \mu^2.
\eea  
Alternatively, using \eqref{eq:lambdaB2}, we can get a more compact expression by letting $\rho'=\mathbf{4}_{\pm}$ to be either of the two $4$-dimensional complex conjugate irreducible representations of $\mathrm{Spin}(6)=\mathrm{SU}(4)$. Since $\wedge^i\rho'=\rho_{i}$ in this case, we obtain right off the bat from \eqref{eq:lambdaB2} that
\beq 
\widetilde{\mathsf{T}}_{B_2,\mathbf{4}_{\pm};u}(\mu,\nu) =
1-(u_1+\nu) \mu+ u_2 \mu^2-(u_1-\nu) \mu^3+ \mu^4.
\label{eq:twistedB2curve}
\eeq 

\end{example}

\begin{example}
As a slightly more complicated example, consider $\big(G_2^{(1)}\big)^\vee = D_4^{(3)}$. 
In this case the  triality symmetry of the simple Lie algebra $\cA(\mathfrak{g}_2)=\mathfrak{d}_4$ identifies the vertices $v_1 \leftrightarrow v_3 \leftrightarrow v_4$, corresponding to the highest weights $[1000]$, $[0010]$ and $[0001]$ of the three eight-dimensional irreducible representations of $\mathrm{Spin}(8)$, whereas $v_2$, corresponding to the adjoint representation, is left fixed; see \cref{fig:dynkG2}. In particular, we have that $\mathrm{Eff}(\mathfrak{g}_2)=\mathsf{o}^{(3)}_1=\{v_1,v_2, v_4\}$. \medskip

\begin{figure}[h]
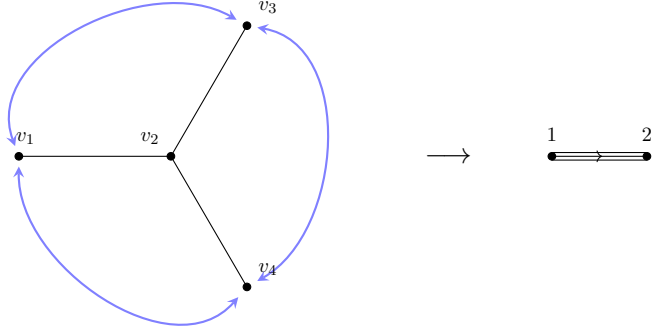

\centering 
\dynkin[%
edge length=2cm,
labels*={~~v_1,v_2,v_3,v_4},
involution/.style={blue!50,stealth-stealth,thick},
involutions={[in=120,out=80,relative]13;[in=120,out=80,relative]34;[in=120,out=80,relative]41}]D4
$\qquad \longrightarrow \qquad$
\dynkin[%
edge length=1.25cm,
labels*={1,2},
]G2
\caption{The Dynkin diagram of $G_2$ via $D_4$-folding.}
\label{fig:dynkG2}
\end{figure}

Fix $\rho'=\rho_{[1000]}$. Then the twisted Cartan--Weyl generators are given in this representation by \cite[Sec.~21.13]{MR1993935}
\bea
e_1^\vee &=&
\varepsilon_{32}-\varepsilon_{76} \nn \\
e_2^\vee &=&
\varepsilon_{21}+\varepsilon_{43}+\varepsilon_{53}-\varepsilon_{64}-\varepsilon_{65}-\varepsilon_{87} \nn \\
e_0^\vee &=& \varepsilon_{14}-\varepsilon_{58} +\omega_3 \left(\varepsilon_{15}-\varepsilon_{48}\right)+ \omega_3^2\left(\varepsilon_{26}-\varepsilon_{37}\right)
\eea 
with again $e_{\bar i}^\vee=(e_i^\vee)^T$, $h_i = [e_i, e_{\bar i}]$, $i=0,1,2$, and where as before $\omega_3=\re^{2\pi\ri/3}$. Plugging this into \eqref{eq:laxtw} it is straightforward to compute the spectral dependence of the exterior traces $\Tr_{\wedge^i \rho} L^\vee_{G_2}$, whose explicit expression we omit here. Using the following relations in the representation ring $\mathrm{Rep}(D_4)$,
\beq 
\wedge^2\rho = \rho_{[0100]}, \quad 
\wedge^3\rho = \rho_{[0010]} \otimes \rho_{[0001]}, \quad \wedge^4\rho = \rho_{[020]} \oplus \rho_{[002]}, 
\eeq 
and 
\bea 
\rho_{[0020]} &=& \rho_{[0010]}\otimes \rho_{[0010]}\oplus \rho_{[1000]},\nn \\
\rho_{[0020]} &=& \rho_{[0010]}\otimes \rho_{[0001]}\oplus \rho_{[1000]}
\eea
it is immediate to verify that the $\lambda$-dependence of the exterior traces $\Tr_{\wedge^i \rho} L^\vee_{G_2}$ is induced by a shift of the fundamental traces as in \eqref{eq:lambdatw}. The reduced spectral polynomial then reads
\bea
\widetilde{T}_{G_2; \rho_1;u}(\mu,\nu) &=& (\mu-1)^2 \left[u_2 (\mu+1)^2 \mu^2-u_1^2 \mu^3-u_1 \left(\mu^5+\mu\right)+\sum_{i=0}^6 \mu^i\right]\nn \\
&+& \nu(\mu-1)^2 \mu \left(\mu^4+2 \mu^3-u_1
   \mu^2+\mu^2+2 \mu+1\right)-\nu^2 \mu^3 \left(\mu^2+\mu+1\right) \nn \\ &+& \frac{3 (\mu+1)^2 \mu^3}{u_0^2}.
\eea 

\end{example}

\section{Picard--Fuchs equations: the simply-laced case}
\label{sec:ade}

\subsection{Extended affine Weyl groups and Frobenius manifolds}
\label{sec:Frobenius}

We start by providing a very brief account of the construction of \cite{MR1606165} of a semi-simple Frobenius manifold on the space of regular orbits of an extended affine Weyl group, referring the reader to  \cite{MR1606165,Brini:2017gfi,Brini:2021pix} for a more extended treatment. 
Let $\mathfrak{g}$ be a complex simple Lie algebra of rank $r$, and write $\mathfrak{h}$ and  $\mathcal{W}_\mathfrak{g}$ for, respectively, its Cartan subalgebra and the Weyl group. Let $\bar k \in \{1, \dots, r\}$ be a label marking the highest weight $\omega_{\bar k}$ corresponding to any of the highest-dimensional fundamental representations of $\mathfrak{g}$. The action of $\mathcal{W}_\mathfrak{g}$ on $\mathfrak{h}$ admits an affine extension $\widetilde{\mathcal{W}}_{\mathfrak{g}, \omega_{\bar k}} \simeq \mathcal{W}_\mathfrak{g} \rtimes \Lambda_\mathfrak{g}^\vee \rtimes \bbZ$ on $\mathfrak{h} \times \bbC$,
\begin{align}
    \widetilde{\mathcal{W}}_{\mathfrak{g}, \omega_{\bar k}} \times \mathfrak{h} \times \mathbb{C} & \rightarrow \mathfrak{h} \times \mathbb{C}\\
    ((w, \alpha^\vee, n), (h, \xi)) & \mapsto (w(h) + \alpha^\vee + n \omega_{\bar k}, \xi - n).
\end{align}
The GIT quotient
\begin{equation}
M^{\rm trig}_{\mathfrak{g}} \coloneqq (\mathfrak{h}^{\text{reg}} \times \mathbb{C})// \widetilde{\mathcal{W}}_{\RR}  = \text{Spec}(\mathcal{O}_{\mathfrak{h} \times \mathbb{C}}(\mathfrak{h}^{\text{reg}} \times \mathbb{C}))^{\widetilde{\mathcal{W}}_{\mathfrak{g}}} \cong \mathcal{T}^{\text{reg}}/\cW_{\mathfrak{g}} \times \mathbb{C}^*, 
   \label{Def:DZmaniLG}
\end{equation}
is isomorphic as a complex manifold to the trivial one-dimensional family $\cM_\mathsf{C}(\cG) \times \bbC^\star$, parametrised by
$\xi:=\log u_0 =-2h_{\mathfrak{g}}\log R_5 \Lambda_{\rm QCD} \in \bbP^1$
of classical vacua with maximally broken gauge symmetry of the pure $\cN=1$ gauge theory on $\bbR^4 \times S_{R_5}^1$ with gauge group the real compact form $\cG$ of $\exp(\mathfrak{g})$.
In \eqref{Def:DZmaniLG}, $\mathfrak{h}^{\text{reg}}$ is the set of regular orbits,  and $\mathcal{T}^{\text{reg}} = \text{exp}(2\pi \ri \mathfrak{h}^{\text{reg}})$ is its image under the exponential map to the maximal torus $\mathcal{T}$.  

In \cite{MR1606165}, the authors construct a canonical, semisimple Frobenius manifold structure on $M_\mathfrak{g}^{\rm trig}$. Their construction asserts that there exist a chart $\{t_i(h,\xi)\}_{i=0}^r$ on $M_\mathfrak{g}^{\rm trig}$ and a solution $F_{\mathfrak{g}}(t_0, \dots, t_r)$ of the WDVV equation such that: %
\ben
\item $F_{\mathfrak{g}} \in \mathbb{Q}[t_0, \dots, t_r][\re^{t_0}]$;
\item $\eta_{ij}:=\de^3_{t_r t_i t_j} F_{\mathfrak{g}}$ is a constant and non-degenerate matrix;
\item $E(F) = 2 F$, with 
\beq E\coloneqq \sum_{j=1}^{r} \frac{\mathfrak{q}_j}{\mathfrak{q}_{\bar k}}t_j \partial_j + \frac{1}{\mathfrak{q}_{\bar k}} \partial_{0},
\label{eq:Evec}
\eeq
up to quadratic terms.
\een
Here, $\mathfrak{q}_i \coloneqq \bra \omega_i, \omega_{\bar k}\ket$ are the fundamental Coxeter degrees of $\mathfrak{g}$.

\subsection{Seiberg--Witten periods as odd periods}

The Frobenius manifolds $M^{\rm trig}_\mathfrak{g}$ are a trigonometric version of the polynomial Frobenius manifolds $M^{\rm pol}_\mathfrak{g}$ \cite{Dubrovin:1993nt,Dubrovin:1994hc} on quotients of the reflection representation of ordinary Weyl groups. For $\mathfrak{g}=\mathfrak{ade}$ these are isomorphic to the chiral rings\footnote{And for $\mathfrak{g}=\mathfrak{bcfg}$, the invariant subrings obtained by Dynkin folding \cite{Zuber:1993vm}.} of the twisted, massively perturbed two-dimensional minimal $\cN=2$ SCFTs  with central charge $d=c/3<1$ \cite{Dijkgraaf:1990dj} of type ADE. It was proposed in \cite{Eguchi:1996nh,Ito:1997ur,Ito:1997zq,MR2070050} that the SW periods of four-dimensional $\cN=2$ super Yang--Mills theories with simply-laced compact gauge group $\cG$ are solutions of a Picard--Fuchs system determined by the Frobenius structure on the tensor product $M^{\rm pol}_\mathfrak{g} \otimes QH(\bbP^1)$ ($\mathfrak{g}= \mathrm{Lie}(\cG)_\bbC$) of the polynomial Frobenius manifold of type $\mathfrak{g}$ with the quantum cohomology of the projective line (the chiral ring of the topologically A-twisted $\sigma$-model). The claim is that there exists a change-of-variables $(u_1, \dots, u_r, \Lambda_{\rm QCD}) \to (t_1, \dots, t_r,Q)$ such that the periods satisfy the holonomic system of PDEs \cite[Prop.~5.19]{MR2070050}
\bea
\label{eq:PF4dhom}
\left(\sum_{i=1}^r \mathfrak{q}_i t_i \de_{t_i}-1\right)^2 \Pi &=& 4 h_{\mathfrak{g}}^2 Q \de^2_{t_r} \Pi \\
\label{eq:PF4prod}
\de^2_{t_i t_j} \Pi &=& c_{ij}^k \de^2_{t_r t_k}  \Pi
\eea 
In \eqref{eq:PF4dhom}--\eqref{eq:PF4prod}, the independent variables $\{t_i\}_{i=1}^r$ are flat coordinates for the Saito metric on $M^{\rm pol}_{\mathfrak{g}}$, and are related polynomially to the classical Coulomb order parameters $u_i$ \cite{Dubrovin:1993nt}; 
the variable~$Q$ keeps track of the dependence on the holomorphic scale, and is identified with the coordinate parametrising primary insertions of the K\"ahler class in the topological A-model on $\bbP^1$ \cite{MR2070050}; and finally, in \eqref{eq:PF4prod}, $c_{ij}^k(t)$ are the structure constants of the $\cN=2$ ADE chiral ring. In the language of \cite{MR2070050}, the gauge theory periods are identified with the {\it odd periods} of the Frobenius manifold $M_{\mathfrak{g}}^{\rm pol} \otimes QH(\bbP^1)$ on the ``coordinate cross'' where all dependence on the primary insertions of the $\bbP^1$ theory are discarded. \medskip

Viewing the pure $\cN=2$ KK theory on $\bbR^4 \times S_{R_5}^1$ with ADE gauge group as a 4d theory with $\widetilde{\rm ADE}$ loop group gauge symmetry, as in \cite{Nekrasov:1996cz}, we propose that the same tensor product construction of \cite{Ito:1997ur,MR2070050} applies to the 5d setting upon replacing the type $\mathfrak{g}=\mathfrak{ade}$ polynomial Frobenius manifolds $M_\mathfrak{g}^{\rm pol}$ with their trigonometric, extended affine version $M_\mathfrak{g}^{\rm trig}$. This leads us to postulate that for ADE gauge groups the periods \eqref{eq:specgeom} satisfy the system of PDEs
\begin{subequations}
\begin{empheq}[box=\widefbox]{align}
\label{eq:PF5Eul}
L_E^2 \Pi =& 4 \de^2_{t_r} \Pi \\
\label{eq:PF5prod}
\de^2_{t_i t_j} \Pi =& \mathsf{C}_{ij}^k \de^2_{t_r t_k}  \Pi
\end{empheq}
\end{subequations} 
where now $i=0,\dots, r$, the Euler vector field $E$ is as in \eqref{eq:Evec}, and $\mathsf{C}_{ij}^k(t)$ are now the structure constants $\mathsf{C}_{ij}^k(t)\coloneqq \eta^{kl} \de^3_{t_l t_i t_j} F^{\rm trig}_\mathfrak{g}$ of $M^{\rm trig}_\mathfrak{g}$ in the flat chart $\{t^i\}_{i=0}^r$.\footnote{A remark is in order regarding the type A series, since in that case we have an additional parametric dependence on the five-dimensional Chern--Simons term. By \eqref{eq:lambdadep}, the tensor product with the quantum cohomology of $\bbP^1$, whose LG superpotential has the form $\lambda+1/(u_0\lambda)$, picks out a CS level such that the symmetry $\lambda \leftrightarrow (u_0 \lambda)^{-1}$ is realised at the spectral curve level. This occurs at $k_{\rm CS}=0$ for $\cG=\mathrm{SU}(2N)$ and $k_{\rm CS}=1$ for $\cG=\mathrm{SU}(2N+1)$.}
\medskip

It will be helpful to express \eqref{eq:PF5Eul}--\eqref{eq:PF5prod} in a natural set of coordinates adapted to the weak coupling expansion in the gauge theory. These are the analogue of the locally monodromy invariant coordinates around the large complex structure/maximally unipotent point in Hori--Vafa mirror symmetries, and they are related to the usual Coulomb branch $u$-coordinates as
\beq
z_0 = 1/u_0, \qquad z_i = \prod_{j=1}^r u_j^{-\cK_{ij}(\mathfrak{g})}
\label{eq:utoz}
\eeq 
where $\cK(\mathfrak{g})$ is the Cartan matrix of $\mathfrak{g}$. In these coordinates, \eqref{eq:PF5Eul}--\eqref{eq:PF5prod} read
\begin{subequations}
\begin{empheq}[box=\widefbox]{align}
\label{eq:PF5Eulz}
\left(z_0 \de_{z_0}\right)^2 \Pi =& 4 h_\mathfrak{g}^2 \left( X^{ij}(z) \de^2_{z_i z_j} + Y^i(z) \de_{z_i}\right) \Pi, \\
\label{eq:PF5prodz}
\de^2_{z_i z_j} \Pi =& \left(A_{ij}^{kl}(z) \de^2_{z_k z_l} + B^{k}_{ij} \de_{z_k}\right)  \Pi,
\end{empheq}
\end{subequations} 
where
\beq 
\bary{rclrcl}
X^{ij} &=& \frac{\de z_i}{\de t_r} \frac{\de z_j}{\de t_r},&  Y^{i} &=& \frac{\de^2 z_i}{\de t_r \de t_k} \frac{\de z_k}{\de t_r},\\
A^{kl}_{ij} &=& \frac{\de t_m}{\de z_i}\frac{\de t_n}{\de z_j}\mathsf{C}^p_{mn}  \frac{\de z_k}{\de t_p}\frac{\de z_l}{\de t_r},&
B^{k}_{ij} &=& \frac{\de t_m}{\de z_i}\frac{\de t_n}{\de z_j}\mathsf{C}^p_{mn}  \frac{\de^2 z_k}{\de t_p \de t_r} + \frac{\partial^2 t_m}{\partial z_i\partial z_j}\frac{\partial z_k}{\partial t_m}.
\label{eq:XYAB}
\eary
\eeq

By design, the large complex structure/semi-classical expansion point corresponds in these coordinates to $z=0$: from \eqref{eq:u0} and \eqref{eq:utoz}, sending $\Lambda_{\rm QCD}$ to infinity sets $z_0=0$, and from \eqref{eq:uvar} sending $a_i$ to infinity leads to a damping behaviour of the form %
\beq
z_i \sim \re^{- \sum_{j}\mathcal{K}_{ij}(\mathfrak{g}) a_j} \sim \re^{-\sum_{j}\mathcal{K}_{ij}(\mathfrak{g}) a^{\rm cl}_j}
\eeq 
so that $z_i \to 0$, $i=1 ,\dots, r$ in that limit. The determination of the gauge theory prepotential from \eqref{eq:PF5Eul}--\eqref{eq:PF5prod} proceeds then along the following steps. 

\ben
\item We write down the $t-$chart Picard--Fuchs system \eqref{eq:PF5Eul}--\eqref{eq:PF5prod} from the Frobenius manifold prepotential $F_\mathfrak{g}^{\rm trig}(t)$, recently found for all $\mathfrak{g}$ in \cite{Brini:2017gfi,Brini:2021pix}.
\item We then derive from this the Picard--Fuchs system \eqref{eq:PF5Eulz}--\eqref{eq:PF5prodz} in the $z$-coordinates, by using the expression of the flat coordinates $t^i(z)$ in terms of the basic invariants/classical gauge theory Casimirs found in \cite{Brini:2017gfi,Brini:2021pix}, and using \eqref{eq:utoz}.
\item We finally look for 
solutions to \eqref{eq:PF5Eulz}--\eqref{eq:PF5prodz} in the form
\bea
\Pi &=& \sum_{j,k} a_{jk}  \log z_j \log z_k + \sum_l  \sum_{J \in \frac{1}{|Z(\cG)|}\bbZ^{r+1}} b_{l,J} \log z_l \prod_k z_k^{J_k} \nn \\ & & + \sum_{J \in \frac{1}{|Z(\cG)|} \bbZ^{r+1}} c_{J} \prod_k z_k^{J_k}
\label{eq:ansatz}
\eea
with at worst double-logarithimic singularities around $z=0$. In \eqref{eq:ansatz} we also allow fractional exponents with denominators that divide the order of the center of the group, as the latter coincides with the determinant of the Cartan matrix, and may arise as a consequence of the change-of-variables in \eqref{eq:utoz}. \medskip

We shall find that \eqref{eq:PF5Eulz}--\eqref{eq:PF5prodz} admit a $2r+2$-dimensional solution space of the form \eqref{eq:ansatz}: two of these are always $\log z_0$ and constant solution,
 and that the remaining $2r$ solutions satisfy the special geometry relations \eqref{eq:specgeom}, from which the gauge theory prepotential can be computed.
\een 

We will put this strategy to the test in some of the lowest rank ADE examples in the next section.

\subsection{Examples}

 

\subsubsection{$A_1$}
We start off by illustrating in detail the simplest example of
$\cG=A_1$ at vanishing $\theta$-angle.
In this case, the Frobenius manifold $M_\mathfrak{g}^{\rm trig}$ coincides with the quantum cohomology ring of $\bbP^1$, and its prepotential is simply
%
\begin{equation}
F_{A_1}^{\rm trig} (t_0,t_1)=t_0t_1^2+\re^{2t_0},
\end{equation}
with Euler vector field $E=\de_{t_0}+t_1 \de t_1$ from \eqref{eq:Evec},
while the flat coordinates $(t_0,t_1)$ are related to the classical Coulomb moduli as \cite{Brini:2021pix}
\begin{align}
t_0=\log u_0,\;\;\;\;t_1=u_0 u_1.
\end{align}
From this, it is immediate to verify that the flat metric is anti-diagonal, $\mathsf{C}_{1i}^j=\delta_i^j$, and
\begin{equation}
\mathsf{C}_{0i}^j=
\left(
\begin{array}{cc}
 0 &  \re^{t_0} \\
 1 & 0 \\
\end{array}
\right).\label{A1Jacobi}
\end{equation}
Finally, the semiclassical expansion coordinates  \eqref{eq:utoz} are:
\begin{equation}
z_0=\frac{1}{u_0},\;\;\;\;z_1=\frac{1}{u_1^2}\label{zA1}
\end{equation}
We now have all the ingredients to write the Picard--Fuchs system \eqref{eq:PF5Eulz}--\eqref{eq:PF5prodz}, which reads
\bea
\left(z_0 \de_{z_0}\right)^2 \Pi &=& 8 z_0^2 z_1^2 (3 \de_{z_1} \Pi + 2 z_1 \de^2_{z_1}) \Pi, \nn \\
(z_0\de_{z_0}^2) \Pi &=& 
4 \left[z_1^2(4 z_1-1) \de_{z_1}^2 + z_1(6 z_1-1) \de_{z_1}
+ z_0 z_1 \de_{z_0} \de_{z_1}\Pi\right] .
\label{eq:PFA1}
\eea 
Inserting the ansatz \eqref{eq:ansatz} into \eqref{eq:PFA1}, we find that the solution space of \eqref{eq:PFA1} is a 4-dimensional complex vector space with coordinates $(b_{0,0},c_{0,0},c_{0,1}, c_{0,2})$:
\bea
& & \Pi =
\frac{2c_{0,2} -3 c_{0,1}}{14} \left[ 
   \log ^2\left(z_1\right)-2\log\left(z_0\right) \log \left(z_1\right)\right] \nn \\
   &+&
\frac{\log z_0}{7} \bigg[\left(7 b_{0,0}+z_1 \left(4 c_{0,2}-6 c_{0,1}\right)+z_1^2 \left(6 c_{0,2}-9 c_{0,1}\right)\right)+z_0^2 \left(4 c_{0,2}-6
   c_{0,1}\right) z_1+\dots \bigg]
   \nn \\
   &+& \frac{\log z_1}{14} \bigg[\left(13 c_{0,1}-4 c_{0,2}\right)+z_1 \left(8 c_{0,2}-12 c_{0,1}\right)+z_1^2 \left(12 c_{0,2}-18 c_{0,1}\right)+z_0^2
   \left(8 c_{0,2}-12 c_{0,1}\right) z_1+\dots \bigg]
   \nn \\
   &+&
  \left(c_{0,0}+z_1 c_{0,1}+z_1^2 c_{0,2}\right)+z_0^2 \left(c_{0,1} z_1+\left(\frac{18 c_{0,1}}{7}+\frac{16 c_{0,2}}{7}\right)
   z_1^2\right)+\dots 
\eea 
Setting $c_{0,1}=2/3 c_{0,2}$ we find a constant holomorphic solution at $c_{0,2}=b_{0,0}=0$, and 
two logarithmic solutions
\bea
\Pi_{A_0} &=& \log z_0 \nn \\
\Pi_{A_1} &=& \log z_1+2 z_1+3 z_1^2+2 z_1 z_0^2+ 12 z_1^2 z_0^2+\dots 
\eea
for $(b_{0,0},c_{0,2})=(1,0)$ and $(0,3)$ respectively. From \eqref{eq:u0} and \eqref{eq:utoz}, we identify $\re^{\Pi_{A_0}}=z_0=q_0$, while $\Pi_{A_1}$ equates to minus the scalar vev $a=\bra \varphi \ket$ in \eqref{eq:specgeom}, in terms of which the inverse mirror map reads
\beq 
z_1(q_0, q_1) =q_1-2 \left(q_0^2+1\right) q_1^2+3 \left(q_0^4+1\right) q_1^3-4 \left(q_0^6+q_0^4+q_0^2+1\right) q_1^4+5 \left(q_0^8-5 q_0^4+1\right) q_1^5+\dots
\eeq 
%
Setting instead $c_{0,1}=1$ and $c_{0,2}=1/4$ gives the dual period
\bea 
\Pi_{B_1} &=& \frac{1}{4} \left(\log ^2z_1+2 \log z_0 \log z_1-4 \log z_1 \right)+\log \left(z_0z_1\right) \left(z_1+\frac{3}{2} z_1^2+z_0^2 z_1+ 6 z_0^2z_1^2+\dots\right) \nn \\ &-& z_1 z_0^2+z_1^2 \left(\frac{1}{4} +2  z_0^2\right)+\dots \nn \\
&=& 
\frac{1}{4} \left(\log^2q_1+2 \log q_0\log q_1-4 \log q_1\right)\nn \\ &+& \left(1+q_0^2\right) q_1+\frac{1}{4}
   \left(1+16 q_0^2+q_0^4\right) q_1^2+\frac{1}{9} \left(1+81 q_0^2+81 q_0^4+q_0^6\right) q_1^3 +\nn \\ &+&\frac{1}{16} \left(1+256 q_0^2+1040 q_0^4+256 q_0^6+q_0^8\right)
   q_1^4+O\left(q_1^5\right)
\eea 
Identifying $\Pi_{B_1}$ with the gradient $\de_{a} \cF$ of the gauge theory prepotential, the above calculation retrieves the expression of the latter for the five-dimensional pure $\mathrm{SU}(2)$ theory at vanishing $\theta$-angle \cite{Nekrasov:1996cz}.

\subsubsection{$A_2$}
As a slightly more difficult case, let us consider another special unitary case given by the $\mathrm{SU}(3)$ gauge theory with $k_{\rm CS}=1$.
The prepotential of $M_{\mathfrak{a}_2}^{\rm trig}$ is given by 
\begin{equation}
F(t_0,t_1,t_2)=\re^{\frac32 t_0}+\frac34 t_0t_2^2-\frac{1}{96}t_1^4+\frac14 t_1^2t_2,
\end{equation}
with flat coordinates related to the classical Coulomb moduli as
\begin{align}
(t_0,t_1,t_2)=(\log u_0,u_0^{\frac12} u_1,u_0 u_2).
\end{align}
As before the flat metric is anti-diagonal, with non-trivial structure constants 
\beq
\mathsf{C}_{0i}^j
=\left(
\begin{array}{ccc}
 0 & 0 & 1 \\
 \frac{9}{2} \re^{\frac{3 t_0}{2}} & 0 & 0 \\
 \frac{9}{4} \re^{\frac{3 t_0}{2}} t(1) & \frac{3}{2} \re^{\frac{3 t_0}{2}} & 0 \\
\end{array}
\right)_{ij}, \quad 
\mathsf{C}_{1i}^j
=
\left(
\begin{array}{ccc}
 0 & \frac{1}{3} & 0 \\
 0 & -\frac{t_1}{2} & 1 \\
 \frac{3}{2} \re^{\frac{3 t_0}{2}} & 0 & 0 \\
\end{array}
\right),
\eeq 
and the Euler vector field is $E=\de_{t_0}+t_1/2 \de_{t_1}+t_2 \de_{t_2}$. Finally, the semiclassical coordinates \eqref{eq:utoz} are here given by
\begin{equation}
z_0=\frac{1}{u_0},\;\;\;\;z_1=\frac{u_2}{u_1^2},\;\;\;\;z_2=\frac{u_1}{u_2^2},\label{A2z}
\end{equation}
The strategy in this case follows {\it verbatim} the discussion of the $A_1$ example. As expected, the solution space of \eqref{eq:PF5Eulz}--\eqref{eq:PF5prodz} is 6-dimensional, linearly generated by one constant, three  logarithmic, and two doubly-logarithmic solutions. By way of example, we obtain for the logarithmic solutions
\bea 
\Pi_{A_0} &=& \log z_0, \nn \\
\Pi_{A_1} &=& \log \left(z_1\right)-z_2-8 z_2 z_1^3+\frac{20 z_1^3}{3}-2 z_2 z_1^2+3 z_1^2-\frac{3 z_2^2}{2} +z_2^2 z_1+2 z_1+6 z_0^2 z_2^{4/3} z_1^{8/3}+\dots\nn \\
   \Pi_{A_2} &=& \log \left(z_2\right)-z_1+2 z_2-\frac{10 z_1^3}{3}-3 z_0^2 z_2^{4/3} z_1^{8/3}+z_2 z_1^2-\frac{3 z_1^2}{2}-8 z_2^3 z_1-2 z_2^2 z_1+\dots
\eea
The final result for the prepotential then is
\bea
-8\pi^3 \ri \cF_{A_2} &=&
(2\log^2q_1 +2\log q_2 \log q_1+4 \log^2q_2)\log q_0 \nn \\ &-& \frac{20 \log^3 q_1}{27}-\frac{10}{9} \log q_2
   \log^2 q_1-\frac{8}{9} \log^2 q_2 \log q_1-\frac{16 \log^3 q_2}{27}\nonumber\\
   &+& 2\left(q_1+2q_1 q_2+2q_2\right)+\frac{1}{4}\left(q_1^2+q_1^2 q_2^2+q_2^2))\right)+\dots
\eea 
Once again we have matched this at the first few orders in $q_i$ with the instanton calculation\footnote{For the $\mathrm{SU}(N)$ theory at an arbitrary Chern--Simons level $k_{\rm CS}$, see \cite[Section 3]{Tachikawa:2004ur} for the perturbative prepotential $\cF_{\mathrm{SU}(N)_{k_{\rm CS}}}^{[0]}$ and \cite{Gottsche:2006bm} for the instanton corrections $\cF_{\mathrm{SU}(N)_{k_{\rm CS}}}^{[n]}$ with $n>0$.}  of the gauge theory prepotential at
 Chern--Simons level 1.  We have verified in the same manner that the same applies for higher rank cases as well, such as $\cG=\mathrm{SU}(4)$ at vanishing CS level, and that moreover our expressions agree with the direct period integral calculations of \cite{Brini:2008rh}.
 
 %

\subsubsection{$D_4$}\label{sec:D4}
Let us now move on to uncharted territory and test the Picard--Fuchs construction of \eqref{eq:PF5Eulz}--\eqref{eq:PF5prodz} on the first non-trivial, non-unitary case corresponding to $\cG=\mathrm{Spin}(8)$. While a direct analysis of period integrals is too hard to carry out in this case, our method allows to compute their weak coupling expansion efficiently.
The prepotential, Euler vector field, and flat coordinates for this case are respectively given by
\begin{align}
F^{\rm trig}_{\mathfrak{d}_4}=&\frac{\re^{2
   t_0}}{2}+\re^{t_0} t_1^2+\re^{t_0} t_3^2+\re^{t_0} t_4^2+2 \re^{\frac{t_0}{2}} t_1 t_3 t_4\nonumber\\
   &+\frac{1}{2} t_2 t_1^2+\frac{1}{2} t_2
   t_3^2+\frac{1}{2} t_2 t_4^2-\frac{t_1^4}{48}-\frac{t_3^4}{48}-\frac{t_4^4}{48}+\frac{t_2^3}{6}+\frac{1}{2} t_0 t_2^2,
   \label{eq:prepd4}
\end{align}
\beq 
E= \de_{t_0}+\frac{t_1}{2} \de_{t_1}+t_2 \de_{t_2}+\frac{t_3}{2} \de_{t_3}+\frac{t_4}{2} \de_{t_4},
\eeq 
\begin{equation}
t_0=\log u_0,\;\;\;\;t_1=u_0^{\frac12} u_1,\;\;\;\;t_2=u_0(u_2+2),\;\;\;\;t_3=u_0^{\frac12} u_3,\;\;\;\;t_4=u_0^{\frac12} u_4.
\end{equation}
The expansion coordinates around the large complex structure point in the moduli space read, from \eqref{eq:utoz},
\begin{equation}
z_0=\frac{1}{u_0},\;\;\;\;z_1=\frac{u_2}{u_1^2},\;\;\;\;z_2=\frac{u_1 u_3 u_4}{u_2^2},\;\;\;\;z_3=\frac{u_2}{u_3^2},\;\;\;\;z_4=\frac{u_2}{u_4^2}.
\label{eq:zd4}
\end{equation}
Explicit expressions for the differential system \eqref{eq:PF5Eulz}--\eqref{eq:PF5prodz} and its solutions are omitted here as they are obtained in a straightforward manner from \eqref{eq:prepd4}--\eqref{eq:zd4}; they are available upon request. Once again our strategy can be shown to retrieve the gauge theory prepotential for this case as well, which we have verified up to one-instanton, and up to $\mathcal{O}(z_i^{12})$. \footnote{As for such high rank the calculations become rapidly quite cumbersome, for our purposes it is quicker to match the special geometry and gauge theory expressions  by a partial reverse-engineering of the solutions of \eqref{eq:PF5Eulz}--\eqref{eq:PF5prodz}. We computed the 
single-logarithmic solutions of \eqref{eq:PF5Eulz}--\eqref{eq:PF5prodz} -- i.e. the mirror map --, and then used these to find a conjectural expression for the dual periods $\Pi_{B_i}$ in terms of the semiclassical variables B-model variables $z_i$ by employing the known results for the 1-loop and 1-instanton contributions to the gauge theory prepotential. We then verified explicitly in an expansion around $z=0$ that these are indeed solutions of our differential system \eqref{eq:PF5Eulz}--\eqref{eq:PF5prodz}.}

\subsubsection{$E_6$}
Finally, we will briefly treat here the exceptional case of $\cG=E_6$. From \cite{Brini:2021pix}, the prepotential of $M^{\rm trig}_{\mathfrak{e}_6}$ reads
\bea 
F^{\rm trig}_{\mathfrak{e}_6} &=& -\frac{t_1^6}{3240}-\frac{1}{648} t_4 t_1^4+\frac{1}{12} \re^{\frac{t_0}{3}} t_5 t_1^4+\re^{t_0} t_1^3+\re^{\frac{t_0}{2}} t_6 t_1^3-\frac{1}{216} t_4^2
   t_1^2+\frac{5}{3} \re^{\frac{2 t_0}{3}} t_5^2 t_1^2 \nn \\ &+& \frac{1}{6} \re^{\frac{2 t_0}{3}} t_2 t_1^2-\frac{1}{6} \re^{\frac{t_0}{3}} t_4 t_5 t_1^2+\frac{1}{6}
   \re^{\frac{t_0}{6}} t_5^2 t_6 t_1^2+\frac{1}{6} \re^{\frac{t_0}{6}} t_2 t_6 t_1^2+\frac{1}{12} \re^{\frac{t_0}{3}} t_5^4 t_1+\frac{1}{12} \re^{\frac{t_0}{3}} t_2^2
   t_1\nn \\ &+& \frac{1}{6} \re^{\frac{t_0}{3}} t_2 t_5^2 t_1+3 \re^{\frac{t_0}{3}} t_5 t_6^2 t_1-\frac{1}{3} t_3 t_4 t_1+3 \re^{\frac{4 t_0}{3}} t_5 t_1-\re^{\frac{t_0}{2}} t_4
   t_6 t_1+6 \re^{\frac{5 t_0}{6}} t_5 t_6 t_1 \nn \\ &+& \frac{\re^{2 t_0}}{2}-\frac{t_5^6}{3240}+\frac{1}{648} t_2
   t_5^4-\frac{t_6^4}{16}+\frac{t_2^3}{648}-\frac{t_4^3}{648}+ \re^{t_0} t_5^3+2 \re^{\frac{t_0}{2}} t_6^3\nn \\ &+& \frac{1}{2} t_0 t_3^2-\frac{1}{216} t_2^2 t_5^2-\frac{1}{6}
   \re^{\frac{2 t_0}{3}} t_4 t_5^2+3 \re^{t_0} t_6^2+ \frac{3}{2} t_3 t_6^2-\frac{1}{6} \re^{\frac{2 t_0}{3}} t_2 t_4+\frac{1}{12} \re^{\frac{t_0}{3}} t_4^2 t_5\nn \\ &+& \frac{1}{3}
   t_2 t_3 t_5+\re^{\frac{t_0}{2}} t_5^3 t_6-\frac{1}{6} \re^{\frac{t_0}{6}} t_4 t_5^2 t_6-\frac{1}{6} \re^{\frac{t_0}{6}} t_2 t_4 t_6+\re^{\frac{t_0}{2}} t_2 t_5 t_6,
\eea
with Euler vector field
\beq
E = \de_{t_0}+ \frac{t_1}{3} \de_{t_1}+ \frac{2t_2}{3} \de_{t_2}+ t_3 \de_{t_3}+ \frac{2t_4}{4} \de_{t_4}+ \frac{t_5}{3}\de_{t_5} + \frac{t_6}{2} \de_{t_6}
\eeq 
and flat and $z$-coordinates given respectively by 
\bea
& t_0=\log u_0,\quad t_1=u_0^{1/3} u_1,\quad t_2=u_0^{2/3} \left(u_1^2-6 u_2-12 u_5\right), \nn \\
& t_3=u_0 \left(u_3+2 u_1 u_5+4 u_6+3\right), \quad t_4=u_0^{2/3}
   \left(12 u_1-u_5^2+6 u_4\right),\nn \\ 
   & t_5=u_0^{1/3} u_5, \quad t_6=\sqrt{u_0} \left(u_6+2\right),
\eea
and
\bea
&
z_0=\frac{1}{u_0},\quad z_1=\frac{u_2}{u_1^2},\quad z_2=\frac{u_1 u_3}{u_2^2}, \quad 
z_3=\frac{u_2 u_4 u_6}{u_3^2}, \nn \\ & z_4=\frac{u_3
   u_5}{u_4^2}, \quad z_5=\frac{u_4}{u_5^2},\quad z_6=\frac{u_3}{u_6^2}.
\eea 
As in the previous case of $\cG=D_4$, to deal with the complexity of the system \eqref{eq:PF5Eulz}--\eqref{eq:PF5prodz} it is convenient to focus on the calculation of single-logarithmic solutions from which the mirror map can be constructed, with the  dual periods being recovered as an {\it a posteriori} check. The $z$-chart differential system \eqref{eq:PF5Eulz}--\eqref{eq:PF5prodz} contains now an unmanageably large number of terms spawned by the change-of-variables \eqref{eq:XYAB}, thereby impeding a direct computational approach based on solving by brute force for the coefficients in the ansatz \eqref{eq:ansatz}. We adopt a hybrid method to circumvent the issue by considering instead the $t$-chart Picard--Fuchs operators \eqref{eq:PF5Eul}--\eqref{eq:PF5prod}, whose expressions are significantly simpler, working out their action on a monomial $\prod_{i=0}^r z_i^{m_i}(t)$, finally expressing the result in $z$-coordinates to obtain recursion relations for the coefficients $(a,b,c)$ in the $z$-chart ansatz \eqref{eq:ansatz}. Restricting e.g. to the single-logarithmic solutions ($a_{jk}=0$, $b_{l,J}=\delta_{J0}\delta_{li}$ for $i=0, \dots, r$) we obtain from \eqref{eq:PF5Eul}
\bea
\left(J_0+2\right)^2 S_0^{-2} \prod_{i=1}^6 S_i^{-2 \mathfrak{q}_i} c_J &=& 4 \bigg[J_2^2+\left(-4 J_3+2 J_4+2 J_6-1\right) J_2+J_4^2+4
   J_3^2 \nn \\ &+& J_6^2-J_4+J_3 \left(-4 J_4-4 J_6+2\right)+2 J_4 J_6-J_6\bigg] c_J,
   \label{eq:receul}
\eea 
where $S_i c_{(J_0, \dots, J_i, \dots J_r)} \coloneqq c_{(J_0, \dots,J_i-1, \dots  J_r)}$ is the left-shift in the $i^{\rm th}$ component of the multi-index $J$ in \eqref{eq:ansatz}. The recursions from \eqref{eq:PF5prod} are obtained similarly; the simplest one is obtained for $(i,j)=(4,2)$, where e.g. we obtain
\bea
0 &=& \left(J_1-2 J_2+J_3\right) \left(J_3-2 J_4+J_5\right) c_J-2 \left(J_3-2 J_6+1\right) \left(J_2-2 J_3+J_4+J_6+1\right)
   S_3S_6 c_J \nn \\ &+& 2 \bigg[J_2^2+\left(-4 J_3+2 J_4+2 J_6-13\right) J_2 \nn \\ &+& 4 J_3^2+J_4^2+J_6^2-13 J_4+J_3 \left(-4 J_4-4
   J_6+26\right)+2 J_4 J_6-13 J_6+42\bigg] S_1S^{2}_2 S_3^4 S_4^2 S_5 S_6^2 c_J \nn \\ &-&\bigg[J_2^2+\left(-4 J_3+2 J_4+2 J_6+3\right) J_2+4
   J_3^2+J_4^2+J_6^2+3 J_4+2 J_4 J_6+3 J_6\nn \\ &-& 2 J_3 \left(2 J_4+2 J_6+3\right)+2\bigg] S_3 c_J.
   \label{eq:rec42}
\eea 
The initial data of the recursions are fixed by the coefficients $(b_{0,0}, \dots, b_{r,0})$ in \eqref{eq:ansatz}, leading to an $r+1$-dimensional vector space of solutions of \eqref{eq:PF5Eulz}--\eqref{eq:PF5prodz} with log-singularities as expected: for example the above differential constraint \eqref{eq:PF5prod} for $(i,j)=(4,2)$ sets 
\beq c_{(0,0,0,1,0,0,0)}=-1/2 c_{(0,1,2,4,2,1,2)}=-b_{0,2}+2b_{0,3}-b_{0,4}-b_{0,6},
\eeq 
while \eqref{eq:PF5Eul} sets $c_{(2,4,8,12,8,4,6)}=-b_{0,2}+2b_{0,3}-b_{0,4}-b_{0,6}$. The full set of recursions for all $(i,j)$ in \eqref{eq:PF5prod} and the ensuing reconstruction of the solutions is omitted for reasons of space, and is available upon request.

\section{Picard--Fuchs equations: the non-simply-laced case}
\label{sec:bcfg}

The construction of the previous Section is constrained to apply to ADE gauge groups: although there exists a definition of an analogous canonical Frobenius manifold structure on the quotient of the reflection representation of extended affine Weyl groups associated to any root system \cite{MR2070050}, including non-simply-laced cases, these are not expected to retrieve the gauge theory periods as solutions of the differential system \eqref{eq:PF5Eulz}--\eqref{eq:PF5prodz}.
The reason for this was already hinted at in \cref{sec:SWcurves}:
the mirror theorem of \cite{Brini:2021pix} asserts that the Frobenius manifolds $M_\mathfrak{g}^{\rm trig}$ are isomorphic to certain Hurwitz strata associated to the spectral curves of the relativistic Toda chain associated to the co-extended loop group of $\cG$, which in the non-relativistic limit reduce to the usual spectral curves of the Toda chain associated to the {\it untwisted} affine Kac--Moody algebra $\mathfrak{g}^{(1)}$. For non-simply-laced $\mathfrak{g}$, this is different from the {\it twisted} Toda chain relevant for the gauge theory: to account for this, it would in fact be natural to speculate that if one replaced, in the Landau--Ginzburg construction of
\cite{Brini:2021pix}, the affine relativistic Toda chain with its twisted (Langlands/Montonen--Olive dual) version, then this would yield a ``twisted''  Frobenius manifold $(M_\mathfrak{g}^{\rm trig})^\vee$ from which the gauge theory periods would be retrieved by the same construction of the previous Section. \medskip 

Both expectations turn out to be false. As we shall see, the natural Frobenius metric associated to the spectral curves of \cref{sec:SWtwToda} is either degenerate or not flat; and perhaps more surprisingly, the twisted relativistic Toda spectral curves will turn out to be the {\it wrong} integrable system for the pure gauge theory in five compactified dimensions, away from $R_5\to 0$. As such, a  new strategy is required to treat the non-simply-laced cases along the same lines as the ADE setting. In the next Section we will describe one such strategy, partly inspired by previous studies of associativity equations for the prepotentials of the 4d and the (perturbative) 5d theory, which also included non-simply-laced cases \cite{Bonelli:1996qh,Hoevenaars:2002wj, Marshakov:1996ae, Marshakov:1997ny,hoevenaars2005wdvv}.

\subsection{The Jacobi algebra of a Seiberg--Witten curve}\label{sec:Jacobi alg}

By the mirror theorem of \cite{Brini:2021pix}, the structure constants $\mathsf{C}_{ij}^k$ in the previous section can be read off from the spectral curve of the type-$\cG^{(1)}$ relativistic Toda chain, $\mathsf{R}_{\cG,\rho';u}(\mu,\lambda)$, as follows. Let $\nu: \overline{C}^{\rm red}_{\cG;u} \to \bbP^1$ be the Cartesian projection to the $\nu$-axis from the perturbative spectral curve $\overline{C}^{\rm red}_{\cG;u}\coloneqq \overline{\{\widetilde{\mathsf{R}}_{\cG,\rho';u}(\mu, \nu/u_0) =0\}}$ for a choice of non-trivial irreducible representation $\rho'$: as the below is independent of the particular choice, we will henceforth suppress $\rho'$ from our notation. Then, from \cite[Thm.~3.1]{Brini:2021pix}, we have that
\beq 
\de_{t_i} \nu(\mu) ~ \de_{t_j} \nu(\mu) = \mathsf{C}_{ij}^k \de_{t_k} \nu(\mu) \de_{t_r} \nu(\mu) + \mathsf{D}_{ij}(\mu) \de_\mu \nu(\mu)
\label{eq:LGalg}
\eeq 
for some meromorphic function $\mathsf{D}_{ij}(\mu)$, and where the $t_i$-derivatives are taken at constant $\mu$. This states that $M_\mathfrak{g}^{\rm trig}$ is isomorphic, as a holomorphic family of commutative rings, to the Jacobi algebra of the meromorphic projection $\nu$.
By the implicit function theorem, we can then rewrite \eqref{eq:LGalg} as
\beq 
\de_{t_i} \widetilde{\mathsf{R}}_{\cG;u} ~ \de_{t_j} \widetilde{\mathsf{R}}_{\cG;u} = \mathsf{C}_{ij}^k~ \de_{t_k}  \widetilde{\mathsf{R}}_{\cG;u}~ \de_{t_{ r}}\widetilde{\mathsf{R}}_{\cG;u} ~ \mod (\widetilde{\mathsf{R}}_{\cG;u}, \de_\mu \widetilde{\mathsf{R}}_{\cG;u})
\label{eq:LGalgpol}
\eeq 
This can be phrased more poignantly as follows. Let $\widetilde{V}_{(i)}(\widetilde{\mathsf{R}}_{\cG;u}) := \de_{t_i} \widetilde{\mathsf{R}}_{\cG;u}/\de_{t_r} \widetilde{\mathsf{R}}_{\cG;u}$. Then for all $u \in \bbC^{r+1}$, the tangent space $T_{t(u)} M_\mathfrak{g}^{\rm trig}$ is isomorphic as a commutative, associative unital algebra to $$\cV(\widetilde{\mathsf{R}}_{\cG;u}) \coloneqq \mathrm{span}_\bbC \{\widetilde{V}_{(i)}\}_{i=0}^r.$$ The latter is obviously a vector subspace of the quotient ring $\bbC[\re^{t_0}, t_1 \dots, t_r][\lambda,\nu]/\cI(\widetilde{\mathsf{R}}_{\cG;u})$, where
$$\cI(\widetilde{\mathsf{R}}_{\cG;u}):=\bra \widetilde{\mathsf{R}}_{\cG;u}(\mu, \nu/u_0), \de_\mu \widetilde{\mathsf{R}}_{\cG;u}(\mu, \nu/u_0)\ket,$$
and \eqref{eq:LGalgpol} states that it is actually a subalgebra, closed under product. 
%
The structure constants $\mathsf{C}_{ij}^k$ (although not the Frobenius metric yet) of $M_\mathfrak{g}^{\rm trig}$ can be obtained from
 \begin{equation}
     \widetilde{V}_{(i)}(\widetilde{\mathsf{R}}_{\cG;u})\widetilde{V}_{(j)}(\widetilde{\mathsf{R}}_{\cG;u})={\mathsf{C}}_{ij}^{k}(t) \widetilde{V}_{(k)}(\widetilde{\mathsf{R}}_{\cG;u})\;\;\;\;{\rm mod} \;\;\;\;\cI(\widetilde{\mathsf{R}}_{\cG;u}),\label{eq:Valg}
 \end{equation}
 in a purely algebraic way, given the knowledge of  $\widetilde{\mathsf{R}}_{\cG;u}$ and the unit vector $e=\widetilde{V}_{(r)}(\widetilde{R}_{\cG;u})$. 
 \medskip
 
How does this help in dealing with non-simply-laced cases? In all instances considered in \cref{sec:SWcurves}, including non-simply-laced gauge groups, the construction of SW curves for the gauge theory leads to a spectral polynomial of the form $\mathsf{P}_{\cG;u}(\mu, \lambda) = \widetilde{\mathsf{P}}_{\cG;u}(\mu,\lambda+q_0/\lambda)$ for some $\widetilde{\mathsf{P}}_{\cG;u} \in \bbZ[\mu, \nu; u_0, \dots, u_r]$, exactly as for the untwisted relativistic Toda polynomials relevant for simply-laced $\cG$ in the discussion above: we have $\mathsf{P}_{\cG;u} = \mathsf{Q}_{\cG;u}$ in the M-theory engineering curves \eqref{eq:QAn}--\eqref{eq:QDn}, and $\mathsf{P}_{\cG;u} = \mathsf{R}_{\cG;u}$  (resp. $\mathsf{P}_{\cG;u} = \mathsf{T}_{\cG;u}$) for the untwisted (resp. twisted) relativistic Toda curves of \eqref{eq:polT}. It is then natural to look at the possibility of replacing the input polynomial $\widetilde{\mathsf{R}}_{\cG;u}$ in \eqref{eq:Valg} with the Hanany--Witten/relativistic Toda reduced characteristic polynomials $\widetilde{\mathsf{Q}}_{\cG;u}$/$\widetilde{\mathsf{T}}_{\cG;u}$, and construct an associated Frobenius manifold out of them. \medskip

Indeed, as we shall show in \cref{claim:ideal} below, for all the curves of \cref{sec:SWcurves} we will be able to associate a holomorphic family of commutative and associative algebras via \eqref{eq:Valg}, and whose existence is itself non-trivial. But these won't come in general from a Frobenius manifold:   
when $\mathsf{P}_{\cG;u}\neq \mathsf{R}_{\cG;u}$:
one can check that the construction of \cite{Dubrovin:1992dz,Brini:2021pix} applied to the the curves in \cref{sec:SWMth,sec:SWtwToda} for non-simply-laced $\cG$ gives rise to a Frobenius metric that is either degenerate (for $\mathsf{Q}_{B_r;u}$ and $\mathsf{T}_{\cG;u}$) or  curved (for $\mathsf{Q}_{C_r;u}$), so there is no clear notion of a privileged ``flat" coordinate system where something like \eqref{eq:PF5Eul}-\eqref{eq:PF5prod} could hold. 
These difficulties can however be side-stepped by imposing the contraints of rigid special K\"ahler geometry: we condense this in the following list of statements.

\begin{claim}\label{claim:ideal}
Let $\widetilde{\mathsf{P}}_{\cG;u}$ be equal to one of $\widetilde{\mathsf{R}}_{\cG;u}$, $\widetilde{\mathsf{Q}}_{\cG;u}$ or $\widetilde{\mathsf{T}}_{\cG;u}$. Up to affine-linear transformations, there exists a unique chart $\{t^i(u)\}_{i=0}^r$
such that 
\ben
\item 
there exist holomorphic functions $\mathsf{C}_{ij}^k(t)$ s.t.
%
%
\begin{align}
    \widetilde{V}_{(i)}(\widetilde{\mathsf{P}}_{\cG;u})\widetilde{V}_{(j)}(\widetilde{\mathsf{P}}_{\cG;u})={\mathsf{C}}_{ij}^{k}(t) \widetilde{V}_{(k)}(\widetilde{\mathsf{P}}_{\cG;u})\;\;\;\;{\rm mod} \;\;\;\;\cI(\widetilde{\mathsf{P}}_{\cG;u})\label{eq:Valg2}
\end{align}

\item the periods associated to the SW curve $\{\mathsf{P}_{\cG;u}(\mu,\lambda)=0\}$ satisfy the system of $2^{\rm nd}$ order PDEs
%
\begin{empheq}[box=\widefbox]{align}
\label{eq:PF5prodgen}
\de^2_{t_i t_j} \Pi =& \mathsf{C}_{ij}^k(t) \de^2_{t_r t_k}  \Pi
\end{empheq}
%
\een 
\vspace{-.75cm}
The coordinates $t_i(u)$ are uniquely fixed by imposing that the inverse $u_i(t)$ is a collection of trigonometric polynomials,
\beq
u_0=\re^{t_0}, \quad u_i \in  \bbC[\re^{-t_0}; t_1, \dots, t_r],
\label{eq:ttougen}
\eeq 
and that single-logarithmic singular solutions of \eqref{eq:PF5prodgen} exist at the point of large complex structure.
\label{claim:coord}
\end{claim}

Our claim therefore postulates the existence of a differential ideal annihilating the periods of the SW curve, which is specified entirely by the product structure on $\cV(\widetilde P_{\cG;u})$ and by the
choice of a canonical system of coordinates $t^i(u)$. The latter behaves as a surrogate of the flat coordinates of the previous section, in spite of the lack of a flat non-degenerate Frobenius metric.

As we did for the ADE gauge groups, we will subject \cref{claim:coord} above to a variety of tests, including comparisons with the gauge theory calculations from \cref{sec:blowup} and direct period integral calculations from the spectral curve, giving in turn an effective method to compute the periods in \eqref{eq:PF5prodgen}--\eqref{eq:PF5Eulgen} around the large complex structure limit. We will proceed, accordingly, in three steps.

\begin{description}

\item[Step 1: the canonical ring structure.]
In analogy with the ADE cases, we assume that some of the special coordinates $t_i$ are determined by the same residue formula as \cite[Lemma.~4.1]{Brini:2021pix} which is a 
specialisation of the residue formulas of \cite[Lecture~5]{Dubrovin:1994hc} for the flat coordinates of a Hurwitz Frobenius manifold. For the examples below, we will indeed find that all $t_i$ (but perhaps $t_r$) are determined by the residue computation. Again similarly to the simply-laced setting, we further impose that $t_r=u_0 (u_r+f(u_1,...u_{r-1}))$ where $f$ is a polynomial in $u_1,...,u_{r-1}$.
We then proceed to verify the first point in \cref{claim:coord} by an algebraic construction of the ring structure on $\cV(\widetilde{\mathsf{P}}_{\cG;u})$. To do this we first compute a reduced Gr\"obner basis $\mathrm{GB}(\widetilde{\mathsf{P}}_{\cG;u})$ for the ideal $\cI(\widetilde{\mathsf{P}}_{\cG;u})$,
and then use multi-variate polynomial division 
to compute the reduction of \eqref{eq:Valg2} w.r.t. $\mathrm{GB}(\widetilde{\mathsf{P}}_{\cG;u})$. Since the latter is a Gr\"obner basis, the reduction is zero iff the above expression is in the ideal $\cI(\widetilde{\mathsf{P}}_{\cG;u})$, and imposing the vanishing of the remainder of the division  gives an {\it a priori} highly overconstrained inhomogeneous linear system for the indeterminates $\mathsf{C}_{ij}^{k}$. While for a general family of plane curves the system would admit no solutions, we will always find that, remarkably, for all the spectral polynomials $\widetilde{\mathsf{P}}_{\cG;u}$ in \cref{sec:SWcurves} a unique solution exists.

\item[Step 2: the special coordinates.]  We are now in a position to write down the differential system \eqref{eq:PF5prodgen} in a quasi-polynomial $t$-chart. At this point we still have a parametric dependence on the coefficients of the polynomial $f(u_1,...,u_{r-1})$, which we fix as follows. 
We first write 
\eqref{eq:PF5prodgen} 
in the $z$-chart defined by \eqref{eq:utoz}, and
look for a condition such that solutions in the form \eqref{eq:ansatz} exist. Remarkably, all coefficients 
of $f(u_1,...,u_{r-1})$  
turn out to be fixed this way, by just looking at the classical piece of the solutions, i.e., their limit as $z_0\to 0$.

\item[Step 3: taming the solutions.] We still lack the analogue of the quasi-homogeneity condition \eqref{eq:PF5Eul} at this stage, which entails that the solution space of \eqref{eq:PF5prodgen} will be {\it a priori} of much higher dimension than the expected $2r+2$.
It turns out that this freedom can be fixed entirely\footnote{With the one exception of the $\mathrm{Sp}(1)$ curves, where a small remaining ambiguity will need additional input to be fixed.}, in either of two ways. For the spectral curves with $\cG=B_r$, associated either to the twisted affine relativistic Toda chain or the M-theory engineering of \cite{Brandhuber:1997ua}, we find that there exist complex numbers $\alpha$, $\beta \in \bbC$, and $l\in \{0,\dots, r\}$ such that

%
%
\begin{empheq}[box=\widefbox]{align}
\label{eq:PF5Eulgen}
\left(t_0+\sum_{i=1}^r \mathfrak{q}_i^\vee t_i \de_{t_i}\right)^2 \Pi =&  \alpha \re^{\beta t_0}\de^2_{t_l t_r} \Pi.
\end{empheq}
%
 In the quasi-homogeneity equation \eqref{eq:PF5Eulgen}, the Euler vector field is specified in terms of the dual Coxeter exponents of $\cG$ (i.e. the $\alpha$-basis coefficients of the highest short root), and we slightly generalise \eqref{eq:PF5Eulgen} by admitting a more general second derivative term in the r.h.s., where possibly $t_l\neq t_r$.\footnote{Although we believe this can be reduced to the form appearing in \eqref{eq:PF5Eulgen}, i.e. with $t_l=t_r$, as an equation on cohomology (the periods), it will be easier for us to consider the generalised, and equivalent equation \eqref{eq:PF5Eulgen} by seeing this as following from a stronger constraint at the level of forms (the SW differential).} For $\cG\neq B_r$, we show that the imposition of the existence of a prepotential whose gradient returns the double-log solutions, $\de_{\Pi_{A_i}}\cF = \Pi_{B_i}$, and whose analytic structure at weak coupling has the form 
 \beq 
 \cF_\cG(q_0,\dots, q_r) = C_3(\log q_0, \dots, \log q_r) + \sum_{I \in (\bbN)^{r+1}} a_I \prod_j q_j^{I_j},\label{eq:Fconstraint}
 \eeq 
allows to fix uniquely all remaining ambiguities.
\end{description}

Having outlined our computational strategy, we will now verify the above proposal in several examples.


\subsection{$B_2$}\label{sec:B2}
From \eqref{eq:QBn}, the perturbative spectral curve for $\cG=B_2=\mathrm{Spin}(5)$ reads
%
\begin{equation}
\widetilde{\mathsf{Q}}_{B_2}(\mu,\nu;u)=\left(\mu^2-1\right) \left(\mu^2+1\right)^2 \mu \,\nu +u_2 \mu^4+u_1 \left(\mu^6+\mu^2\right)+\mu^8+1.
\label{B2curve}
\end{equation}

As far as \cref{claim:coord} is concerned, we find that for this example, and indeed for all $B_r$ cases we have checked (up to $r=4$), the privileged coordinate set $t_i(u)$ including $t_r$ can be found by the same residue calculation of \cite[Lemma~4.1]{Brini:2021pix} applied to the perturbative SW curve \eqref{B2curve}. We thus get
\begin{equation}
t_0=\log u_0,\;\;\;\;t_1=u_0^{1/2}\sqrt{2 - 2 u_1 + u_2},\;\;\;\;t_2=u_0(2+2u_1+u_2).
\end{equation}
with inverse
\beq 
u_0= \re^{t_0},~u_1 = -\frac{1}{4} \re^{-t_0} \left(t_1^2-t_2\right),~u_2= \frac{1}{2} \re^{-t_0} \left(t_1^2-4 \re^{t_0}+t_2\right).
\eeq 
A reduced Gr\"obner basis for the discriminant ideal $\cI(\widetilde{\mathsf{Q}}_{B_2;u})$ can be computed using Buchberger's algorithm with respect to the lexicographic monomial ordering $\mu \prec \nu$. We obtain $\mathrm{GB}(\widetilde{\mathsf{Q}}_{B_2;u}) = \{P(\mu), Q(\mu,\nu)\}$, with
\bea 
P(\mu) &=& \left(\mu ^2+1\right)^4 \left(4 \left(\mu ^2-1\right)^2 e^{t_0}-\mu ^2 t_2\right)+\mu ^2 \left(\mu ^2-1\right)^2 \left(\mu ^4-6 \mu ^2+1\right) t_1^2, \nn \\
   Q(\mu,\nu) &=& \mu  \Bigg[\left(39 \mu ^8-313 \mu ^6+557 \mu ^4-347 \mu ^2+64\right) t_1^2\nn \\ &+& 4 (\mu^2 -1) \left(39 \mu ^8+116 \mu ^6+78 \mu ^4-68 \mu ^2-101\right)
   e^{t_0} \nn \\ &-& \left(39 \mu ^8+155 \mu ^6+233 \mu ^4+165 \mu ^2+64\right) t_2\Bigg]+128\nu. \quad
   \eea 
Reducing \eqref{eq:Valg2} to the ideal $\cI(\widetilde{\mathsf{Q}}_{B_2;u})=\bra \mathrm{GB}(\widetilde{\mathsf{Q}}_{B_2;u}) \ket$ w.r.t. to the reduced basis $\mathrm{GB}(\widetilde{\mathsf{Q}}_{B_2;u})$ and solving for $\mathsf{C}_{ij}^k$, we obtain, for
the structure constants of \eqref{eq:PF5prodgen},
\begin{equation}
\mathsf{C}_{0i}^j=\left(
\begin{array}{ccc}
 -t_1^2-16 \re^{t_0}+t_2 & -16 \re^{t_0} t_1 & 16 \re^{t_0} t_2 \\
 -2 t_1 & -16 \re^{t_0} & 0 \\
 1 & 0 & 0 \\
\end{array}
\right)_i^j,
\end{equation}
\begin{equation}
\mathsf{C}_{1i}^j=\left(
\begin{array}{ccc}
 -2 t_1 & -16 \re^{t_0} & 0 \\
 -2 & -t_1 & 2 t_2 \\
 0 & 1 & 0 \\
\end{array}
\right)_i^j.
\end{equation}
Finally, let us show how to derive the quasi-homogeneity condition \eqref{eq:PF5Eulgen}. We will directly impose that \eqref{eq:PF5Eulgen} holds as a constraint on the Seiberg--Witten differential $\log \mu \rd\log \lambda$ for some $\alpha$, $\beta$ and $l$. The full SW curve $\overline{\cC}_{B_r;u}$ is hyperelliptic, with a degree two map $\mu:\overline{\cC}_{\cG;u} \to \bbP^1$ realising it as a branched cover of the complex line. In particular there is a subset of homology cycles $\{\gamma_1, ..., \gamma_{g(\overline{\cC}_{B_r;u})}\}$ in $\overline{\cC}_{B_r;u}$ such that
\beq 
\Pi_{\gamma_i} \coloneqq \oint_{\gamma_i} \log \mu \rd\log \lambda = \int_{l_i} \left(\log \lambda_+ -\log \lambda_-\right) \rd \log \nu
\label{eq:hyperPi}
\eeq 
for chains $l_i=\mu(\gamma_i) \subset \bbP^1$, where an integration by parts has been performed. Let now $\mathfrak{D}=\sum_{ij} \mathsf{a}_{ij}(u) \de^2_{u_i u_j} + \sum_i \mathsf{b}_i \de_{u_i}$ be a second order differential operator whose symbol $\sigma(\mathfrak{D})$ has vanishing constant term. Then it is easy to show that $\mathfrak{D}(\log \lambda_+(\mu) -\log \lambda_-(\mu)) = P_1/P_2^{3/2}$ for $P_1,P_2 \in \bbC[\mu; u_0, \dots, u_r]$. Imposing that $P_1 \equiv 0$ identically in $\mu$ gives an in principle overconstrained system of equations in the indeterminates $\mathsf{a}$, $\mathsf{b}$. Happily, this can be shown to non-trivially admit a 1-dimensional space of solutions, parametrised by overall rescalings of $\mathfrak{D}$: in flat co-ordinates and at the level of periods, this gives 
\begin{equation}
     \left(\de_{t_0} + \frac12 t_1\de_{t_1} + t_2 \de_{t_2} \right)^2\Pi =-256 \re^{-t_0}\partial_{t_0}\partial_{t_2}\Pi.
\end{equation}
completing the construction of the PF system in \eqref{eq:PF5prodgen} and \eqref{eq:PF5Eulgen}. \medskip 

Armed with this, we proceed as in the analysis for the simply-laced cases. We find three linearly independent single-log solutions in the $z_i$-coordinates \eqref{eq:utoz}, and two double-log solutions, confirming our \cref{claim:coord}. 
 Up to 2-instanton corrections, we find that the perturbative level is unambiguously fixed by \eqref{eq:PF5prodgen}, while order by order in $z_0$, each solution comes with finitely many constants which are equivalently fixed by the quasi-homogeneity equation \eqref{eq:PF5Eulgen} or by imposing imposing the existence of the prepotential. 
We find that the final result,
\bea
-4\pi^3 \ri \cF_{B_2} &=&
\frac{5}{6} \log ^3q_2+\log q_0 \log ^2q_2+\frac{5}{4} \log q_1 \log ^2q_2+\frac{3}{4} \log
   ^2q_1 \log q_2+\log q_0 \log q_1 \log q_2
   \nn \\
   &+& 
q_2 +q_1 q_2+q_1 q_2^2  +\frac{q_2^2}{8} +\frac{q_2^3}{27} +\frac{q_2^4}{64}+\frac{q_1^2 q_2^2}{8}+\dots \nn \\
 &+& 
q_0 \Big( q_1 q_2 +3 q_1^2 q_2 +5 q_1^3 q_2+4 q_1^2q_2^2  +7 q_1^4 q_2+8  q_1^3q_2^2+3  q_1^2q_2^3 +9 q_1^5q_2 +12  q_1^4q_2^2 \nn \\ &+& 9  q_1^3 q_2^3 +11  q_1^6q_2 +16  q_1^5q_2^2 +15 q_1^4q_2^3 +8  q_1^3q_2^4+\dots\Big)
\eea  
is in precise agreement with the IMS, 1-loop, and 1-instanton prepotential, which we have checked up to $\cO(q_i^8)$. 


\subsection{$B_3$}

The reduced characteristic polynomial for the M-theory curve of $\cG=B_3$ is given from \eqref{eq:QBn} as 
\begin{align}
\widetilde{\mathsf{Q}}_{B_3;u}(\mu,\nu)=&(\mu^2-1) \mu^3  \left(\mu^2+1\right)^2 \nu \nonumber\\
&+u_3 \mu^6+u_1 \left(\mu^{10}+\mu^2\right)+u_2
   \left(\mu^8+\mu^4\right)+\mu^{12}+1,
\end{align}
We follow exactly the same procedure as for $\cG=B_2$ to compute the structure constants of the Jacobi algebra \eqref{eq:Valg} and the flat coordinates. We omit details and present the final result for both. For the coordinates, we get
%
\beq
t_0=\log u_0,~ t_1=u_0^{\frac23}(u_1-1),~t_2=u_0^{1/2} (2 u_1-2u_2 + u_3-2)^{1/2}, 
~t_3=u_0(2+2u_1+2u_2+u_3),
\eeq
with polynomial inverses
\beq
u_0= \re^{t_0},~u_1= \re^{-\frac{2 t_0}{3}} t_1+1, ~u_2= \frac{1}{4} \re^{-t_0} t_2^2-\frac{1}{4} \re^{-t_0} t_3+1,~u_3= \frac{1}{2} \re^{-t_0} t_2^2-2 \re^{-\frac{2 t_0}{3}} t_1+\frac{1}{2} \re^{-t_0} t_3-2,
\eeq 
while the structure constants in the $t$-chart read
\begin{equation}
\mathsf{C}_{00}^j={
\left(
\begin{array}{c}
 \frac{4 \re^{\frac{t_0}{3}} t_1 \left(t_1-24 \re^{\frac{t_0}{3}}\right)+3 t_3-3 t_2^2}{9}  \\ \frac{12 \re^{\frac{2 t_0}{3}} t_1^2+288
   \re^{t_0} t_1+3 \left(t_3-t_2^2\right) t_1-432 \re^{\frac{4 t_0}{3}}+\re^{\frac{t_0}{3}} \left(18 \left(5 t_2^2+t_3\right)-8 t_1^3\right)}{27}  \\ \left(16
   \re^{t_0}-\frac{16}{3} \re^{\frac{2 t_0}{3}} t_1\right) t_2 \\ \frac{16 \re^{\frac{2 t_0}{3}}}{3}  
\end{array}
\right)_j
}
\end{equation}
\begin{equation}
\mathsf{C}_{1i}^j=\left(
\begin{array}{cccc}
 \frac{4}{3} \re^{\frac{t_0}{3}} \left(t_1-6 \re^{\frac{t_0}{3}}\right) & \frac{1}{9} \left(-8 \re^{\frac{t_0}{3}} t_1^2-36 \re^{\frac{2 t_0}{3}} t_1+72 \re^{t_0}-3 t_2^2+3
   t_3\right) & -\frac{16}{3} \re^{\frac{2 t_0}{3}} t_2 & \frac{16}{3} \re^{\frac{2 t_0}{3}} t_3 \\
 4 \re^{\frac{t_0}{3}} & -\frac{4}{3} \re^{\frac{t_0}{3}} \left(2 t_1+9 \re^{\frac{t_0}{3}}\right) & 0 & 0 \\
 0 & -2 t_2 & -16 \re^{\frac{2 t_0}{3}} & 0 \\
 0 & 1 & 0 & 0 \\
\end{array}
\right)_i^j
\end{equation}
\begin{equation}
\mathsf{C}_{2i}^j=
\left(
\begin{array}{cccc}
 -2 t_2 & 8 \re^{\frac{t_0}{3}} t_2 & 16 \re^{t_0}-\frac{32}{3} \re^{\frac{2 t_0}{3}} t_1 & 0 \\
 0 & -2 t_2 & -16 \re^{\frac{2 t_0}{3}} & 0 \\
 -6 & 2 \left(t_1+6 \re^{\frac{t_0}{3}}\right) & -t_2 & 2 t_3 \\
 0 & 0 & 1 & 0 \\
\end{array}
\right)_i^j,
\end{equation}
and $\mathsf{C}_{3i}^j=\delta_{i}^j$. As for the $B_2$ case, we also look for a smaller set of differential equations satisfied directly by the SW differential at for fixed $\mu$. It turns out that there is a four-dimensional space of differential operators annihilating \eqref{eq:hyperPi} in this case, and one basis element is in the desired form:
\begin{equation}
    \left(\de_{t_0} + \frac23 t_1\de_{t_1} +\frac12 t_2 \de_{t_2} + t_3 \de_{t_3} \right)^2\Pi=-4096 \re^{-t_0/3}\partial_1\partial_3\Pi.
\end{equation}

Once again the corresponding solutions are found to have the desired asymptotic behaviour at infinity in the Coulomb branch. We verified that the resulting prepotential agrees with the gauge theory calculation up to 1-instanton corrections and up to order $\mathcal{O}(q_i^7)$, $i=1,2,3$.





\subsection{$C_1$ at $\theta=0$}

For symplectic gauge groups, $\cG=C_r$, we have the two different sets of curves \eqref{eq:QCnb} and \eqref{eq:QCny0}--\eqref{eq:QCnypi}, from 
\cite{Brandhuber:1997ua} and \cite{Hayashi:2017btw,Li:2021rqr} respectively. We consider here for the purpose of illustration the rank one case at different theta angles. \medskip

For $\theta=0$, \eqref{eq:QCny0} gives 
\begin{equation}
\widetilde{\mathsf{Q}}_{C_1;u,0}(\mu,\nu/u_0)=\mu^3 \nu+\left(\mu^2-1\right)^2 \left(u_1 \mu+\mu^2+1\right)+\frac{\mu^2 \left(\mu^2+1\right)}{u_0}.
\end{equation}
%
For the coordinates $t^i(u)$ and the $t-$chart structure constants, we find
\begin{equation}
t_0=\log u_0,\;\;\;\;t_1=u_0 u_1,
\end{equation}
\begin{equation}
\mathsf{C}_{0i}^j=\left(
\begin{array}{cc}
 -\frac{2 t_1}{3} & \frac{1}{3} \re^{t_0} \left(4 \re^{t_0}-1\right) \\
 1 & 0 \\
\end{array}
\right),
\end{equation}
and again $\mathsf{C}_{1i}^j=\delta_i^j$.\\

The resulting Picard--Fuchs system is superficially different from the one of $A_1$ in \eqref{A1Jacobi}. Also the absence of an analogue of the quasi-homogeneity condition \eqref{eq:PF5Eulgen} leads to a space of solutions of dimension higher than $2r+2=4$: while for higher rank the imposition of the existence of a prepotential is sufficient to project to a solution space of the correct dimension, the case of rank one is pathological, and leads to a small finite ambiguity at every instanton level. In this case, perturbative parts and 1-instanton corrections are unambiguously fixed, but for higher instanton corrections we find two free parameters at each even order, and none at odd orders, which we can fix by matching with a direct period expansion from the curve. By doing so, we verify that the resulting prepotential is in agreement with the $\mathrm{SU}(2)$ calculation at vanishing $\theta$-angle, which we confirmed in an expansion at weak coupling up to 5-instantons.

\subsection{$C_1$ at ${\theta=\pi}$}
\label{sec:c1pi}
We will jointly consider in this section the curves \eqref{eq:QCnypi} and \eqref{eq:QCnb}.
%
We will show that, while formally different, these surprisingly give rise to the same algebra structure on the discriminant of the perturbative SW curve, and therefore to the same differential system \eqref{eq:PF5prodgen}. As in the previous Section, however, in this rank one situation there will be a finite (two-dimensional) ambiguity at each order of the $q_0$ expansion of the solutions, which as for $\theta=0$ we fix from a direct period calculation from the full SW curve. It turns out that the free parameters thus fixed will be {\it different} according to which curve one picks:  the respective prepotential will also be different, with only the curve \eqref{eq:QCnypi} returning the correct gauge theory prepotential.
\\

The perturbative SW curve  \eqref{eq:QCnb}
of \cite{Brandhuber:1997ua} reads
\begin{equation}
\widetilde{\mathsf{Q}}_{C_1;u}^\flat(\mu,\nu)=\mu^3 \nu+\left(\mu^2-1\right)^2 \left(u_1 \mu-\mu^2-1\right),\label{C1}
\end{equation}
whereas the $C_1^{\theta=0}$ curve in \cite{Hayashi:2017btw} is given by
\begin{equation}
\widetilde{\mathsf{Q}}_{C_1;u,\pi}(\mu,\nu)=\mu^3\nu-\left(\mu^2-1\right)^2 \left(u_1 \mu+\mu^2+1\right)+\frac{2 \mu^3}{u_0}\label{C10}
\end{equation}
Note that, up to rescalings of $\nu$ and $\mu$, the curves differ only by the last term which depends on $u_0$, and is therefore proportional to the gauge theory scale: this implies that the 1-loop prepotentials agree for these two curves. 
It is however less trivial to see whether this equality survives instanton corrections.  In the same special coordinates for $\cG=A_1$ and $\cG=C_1^{\theta=0}$, 
\begin{equation}
t_0=\log u_0,\;\;\;\;t_1=u_0 u_1,
\end{equation}
we find that
the structure constants for both curves indeed surprisingly agree:
\begin{equation}
{\mathsf{C}}_{0i}^j=\left(
\begin{array}{cc}
 -\frac{2 t_1}{3} & \frac{4 \re^{2 t_0}}{3} \\
 1 & 0 \\
\end{array}
\right),\label{C1Jacobi}
\end{equation}
and ${\mathsf{C}}_{1i}^j=\delta_i^j$.  We then solve \eqref{eq:PF5prodgen} in an expansion
in $z_0=1/u_0$, $z_1=1/\sqrt{u_1}$ around $(z_0,z_1)=(0,0)$.

As before, in this rank-1 case the imposition of the existence of a prepotential does not give enough constraints, and the equations \eqref{eq:PF5prodgen} leave a two-dimensional ambiguity at every instanton level which we fix, at any given order in $q_0$, by an expansion of the period integrals to leading order in $z_1=1/u_1$. It turns out that different values are obtained for these free parameters according to whether one picks \eqref{eq:QCnypi} or \eqref{eq:QCnb}: for example, for the non-trivial single-log solution of \eqref{eq:PF5prodgen}, we find
\beq
\Pi_A =
\log z_1+z_1^2+C_1 z_0 z_1^3+\frac{3 z_1^4}{2}+6 C_1 z_0 z_1^5+z_1^6 \left(C_2 z_0^2+\frac{10}{3}\right)+30 C_1 z_0 z_1^7+z_1^8 \left(14 C_2 z_0^2+\frac{35}{4}\right)+\cO\left(z_1^9\right)
\eeq 
where $(C_1,C_2,\dots)=(2,15,\dots)$ for \eqref{eq:QCnypi} and $(C_1, C_2,\dots)=(0,5,\dots)$ for \eqref{eq:QCnb}. These impact the prepotential directly: for the curve \eqref{eq:QCnypi} we find 
\bea 
-4 \pi^3 \ri\cF_{C_1,\theta=\pi} &=& 
\frac{1}{8} \log ^2 q_1\log q_0+\frac{1}{12}\log^3 q_1 + q_1+\frac{q_1^2}{8}+\frac{q_1^3}{27}+\frac{q_1^4}{64}+ \dots \nn \\ &-& \frac{q_0}{2} \left(1+
   3 q_1 +5 q_1^2+ 7 q_1^3+ 9 q_1^4+ 11 q_1^{5}+\dots\right) \nn \\ &+& 
   q_0^2  \left(-\frac{q_1}{16}+ \frac{45}{16} q_1^2 + 16 q_1^3 + \frac{875}{16} q_1^4 + 144 q_1^5+ \dots  \right)
   \label{eq:prepC1pi},
   \eea 
in precise agreement with the genus zero topological string calculation from the engineering geometry $K_{\mathbb{F}_1}$ \cite{Chiang:1999tz}. For the curve \eqref{eq:QCnb} instead we get the correct expression for the 1-loop term, but disagreement is already found at 1-instanton level for all terms, where e.g. the $[q_0^1 q_1^0]$ coefficient in the expansion of the prepotential  is predicted to vanish. This confirms the tension between the results of \cite{Brandhuber:1997ua} on one hand and \cite{Hayashi:2017btw,Li:2021rqr} on the other, ruling out the curve \eqref{eq:QCnb} of \cite{Brandhuber:1997ua} as a candidate Seiberg--Witten geometry for the $\mathrm{Sp}(1)$ theory with no flavours.

\subsection{The relativistic Toda chain on a twisted loop group}
\label{sec:PFToda}

\subsubsection{$B_2$}
From \eqref{eq:twistedB2curve}, the spectral curve of the twisted $B_2$ relativistic Toda chain can be obtained from the reduced characteristic polynomial
\begin{equation}
\widetilde{\mathsf{T}}_{B_2;u}(\mu,\nu)=u_2 \mu^4-u_1 \left(\mu^3+\mu\right)+\mu^4+1+\left(\mu^2-1\right) \mu \nu.\label{twistedB2}
\end{equation}
Note that \eqref{twistedB2} is different from \eqref{B2curve}, even after the natural redefinition $\mu \to \sqrt{\mu}$. In the non-relativistic limit, the full curve $\mathsf{T}_{B_2;u}(\mu,\lambda)=\widetilde{\mathsf{T}}_{B_2;u}(\mu,\lambda +q_0/\lambda)=0$ reduces to the four-dimensional $(B_2^{(1)})^\vee = A_3^{(2)}$ curve of \cite{Martinec:1995by}. \medskip

Even though $t_2$ is not determined by the residue formula of \cite[Lemma~4.1]{Brini:2021pix}, we can still find it by applying the strategy outlined above. Omitting computational details, the special geometry condition on the dual periods gives for the $t$-chart the following expressions:
\begin{equation}
t_0=\log u_0,\;\;\;\;t_1=u_0 u_1,\;\;\;\;t_2=u_0(2+u_1+u_2).
\end{equation}
 The Gr\"obner basis calculation for $\cI\big(\widetilde{\mathsf{T}}_{B_2;u}\big)$ gives then
%
\bea
& & \mathrm{GB}\big(\widetilde{\mathsf{T}}_{B_2;u}\big) =
\Big\{\re^{-t_0} \left(\left(\mu ^2+4 \mu +1\right) \mu ^2 t_1+\left(\mu ^2+1\right) \left(\left(\mu ^2-1\right)^2 \re^{t_0}-\mu ^2 t_2\right)\right), \nn \\ & & \re^{-t_0} \left(2
   \nu +3 \mu ^5 \re^{t_0}-2 \mu ^3 \re^{t_0}-\left(3 \mu ^2+4\right) \mu  t_2+\left(3 \mu ^3+12 \mu ^2+4 \mu +2\right) t_1-\mu  \re^{t_0}\right)\Big\},\quad
\eea
from which the structure constants of the Jacobi algebra are read off as
\begin{equation}
\mathsf{C}_{0i}^j=
\left(
\begin{array}{ccc}
 -t_1-4 \re^{t_0}+t_2 & 4 \re^{t_0} t_1 & 4 \re^{t_0} t_2 \\
 -1 & t_2-t_1 & 3 t_1+t_2 \\
 1 & 0 & 0 \\
\end{array}
\right),
\end{equation}
\begin{equation}
\mathsf{C}_{1i}^j=
\left(
\begin{array}{ccc}
 -1 & t_2-t_1 & 3 t_1+t_2 \\
 \re^{-t_0} & -2 & 3 \\
 0 & 1 & 0 \\
\end{array}
\right).
\end{equation}
As for the $B_2$ curve of \cite{Brandhuber:1997ua} considered in Section~\ref{sec:B2}, since $\overline{\{ \mathsf{T}_{B_2;u}(\mu, \lambda+q_0/\lambda)=0}\}$ is hyperelliptic, we can directly derive a quasi-homogeneity equation satisfied by the Seiberg--Witten differential in the form \eqref{eq:PF5Eulgen}:
\begin{equation}
    L_E^2\Pi=4 \re^{-t_0}\partial_0\partial_2\Pi.
\end{equation}
With the above in place, it is now straightforward to compute the prepotential. Surprisingly, this {\it disagrees} with the gauge theory result, away from the four-dimensional, $R_5\to 0$ limit, already at the 1-loop level. We find:
\bea 
-4\pi^3\ri \cF_{B_2} &=& \frac{\log^3 q_1}{3}+\frac{1}{2} \log q_0 \log^2 q_1+\frac{1}{2}  \log^2 q_1\log q_2+\frac{3}{8} \log q_1\log^2 q_2 \nn \\ 
&+& \frac{1}{2} \log q_0 \log q_1\log q_2 +\frac{\log^3 q_2}{8}+\frac{1}{4} \log q_0 \log^2 q_2
\nn \\
&+& 
\frac{q_2}{4}+\frac{q_2^2}{32}+\frac{q_2^3}{108}+q_1+q_1
   q_2+\frac{ q_1^2}{8}+\frac{ q_1^2q_2}{4}+\frac{ q_1^2q_2^2}{8} + \frac{q_1^3}{27}+\frac{q_1^3 q_2^3}{27}
   +\cdots
\nn \\
&+&
q_0\left(\frac{1}{2} q_1^2 q_2+q_1^3 q_2+\frac{3}{2}  q_1^4q_2+ q_1^3q_2^2+2  q_1^5q_2+2  q_1^4q_2^2 +\frac{5}{2}  q_1^6q_2 +3  q_1^5q_2^2+\frac{3}{2} 
   q_1^4q_2^3+\dots\right)
   \nn \\
&+&
q_0^2\left(\frac{1}{32} q_1^4 q_2^2+\frac{3}{4}  q_1^5q_2^2+\frac{1}{16}  q_1^4q_2^3+\frac{65}{16}  q_1^6q_2^2+ q_1^5q_2^3+\frac{1}{8} 
   q_1^4q_2^4+\frac{55}{4}  q_1^7q_2^2+\frac{99}{16}  q_1^6q_2^3+\dots\right).
   \nn \\
\label{weightedF0B2}
\eea
We see from \eqref{weightedF0B2} that the W-bosons multi-cover contributions to the prepotential take the form $\sum_{\alpha \in \Delta^+(B_2)} 4/(\alpha, \alpha)^2  \mathrm{Li}_3(q^\alpha)$, where contributions of long roots appear weighted with their inverse square length ($=1/4$), instead of appearing with weight one. Note that this is not incompatible with the curve being correct in the four-dimensional limit, as such weighing factor can be reabsorbed when $R_5\to 0$ by a rescaling of the Coulomb moduli and an immaterial quadratic shift of the prepotential. 
It should not come as much of a surprise that the higher order instanton terms also disagree with the gauge theory calculation from the blowup formulas \eqref{1instanton}.

%
\medskip

At this point one might wonder whether the mismatch with the instanton calculation is to be ascribed to a failure of our formalism in this case, rather than a genuine deviation of the twisted Toda spectral curve from the correct description of the low energy theory. A good litmus test would be to contrast \eqref{weightedF0B2} with a direct period integral calculation from the curve: as the discrepancy with the gauge theory prepotential affects the perturbative level as well, this can be probed already in the $q_0\to 0$ limit, where by  \eqref{eq:lambdatw} the spectral curve reduces to the vanishing locus of \eqref{twistedB2}. This is a rational curve, as the $\mu$-projection is unramified and maps it isomorphically to a $\bbP^1$. As a result, the general expression for the triple derivatives of the prepotential boils down to a sum of residues at the divisors of zeroes and poles of $\mu$ and $\nu$:
\bea
\de^3_{\log q_i, \log q_j, \log q_k} \cF_{B_2} &=& \sum_{\rd \nu=0} \frac{q_i q_j q_k\de_{q_i}(\log\mu~\rd\log\nu)\de_{q_j}(\log\mu~\rd\log\nu)\de_{q_k}(\log\mu~\rd\log\nu)}{\rd \log \mu ~\rd \log \nu} \nn \\
&=& \sum_{\nu,\mu\in \{0,\infty\}} \frac{q_i q_j q_k\de_{q_i}(\log\nu~\rd\log\mu)\de_{q_j}(\log\nu~\rd\log\mu)\de_{q_k}(\log\nu~\rd\log\mu)}{\rd \log \mu ~\rd \log \nu}\nn \\
\label{eq:tripleB2}
\eea 
where in going from the first to the second line we have turned the contour around in the $\mu$-plane and picked up poles of the integrand in the complement of $\{\rd \nu(\mu)=0\}$; we have further used the fact that $\{\rd \mu(\nu)=0\}=\emptyset$; and we have finally employed the `thermodynamic identity' $\de_{q_i} \mu \rd\nu=-\de_{q_i} \nu \rd\mu$ \cite[Lecture~5]{Dubrovin:1994hc}. The residues in \eqref{eq:tripleB2} are then immediate to compute as, from \eqref{twistedB2}, they are just residues of the rational function $\nu(\mu)$ at $\mu=0$, $\infty$, and $\re^{\pi\ri k/2}$, $k=0,1,2,3$. This returns the $q_0=0$ part of \eqref{weightedF0B2}, confirming the correctness of the proposed Picard--Fuchs system, and the discrepancy with the gauge theory alongside it.

\subsubsection{$G_2$}
Our last example is the spectral curve geometry of the twisted relativistic Toda chain associated to the $G_2$ root lattice. The reduced characteristic polynomial is given by
\bea
\widetilde{T}_{G_2;u}(\mu,\nu) &=& (\mu-1)^2 \left[u_2 (\mu+1)^2 \mu^2-u_1^2 \mu^3-u_1 \left(\mu^5+\mu\right)+\sum_{i=0}^6 \mu^i\right]\nn \\
&+& \nu(\mu-1)^2 \mu \left(\mu^4+2 \mu^3-u_1
   \mu^2+\mu^2+2 \mu+1\right)-\nu^2 \mu^3 \left(\mu^2+\mu+1\right) \nn \\ &+& \frac{3 (\mu+1)^2 \mu^3}{u_0^2}.
\eea 
The corresponding spectral curve $\{\widetilde{\mathsf{T}}_{G_2, u}(\mu,\nu+q_0/\nu)=\mathsf{T}_{G_2, u}(\mu,\lambda)=0\}$ has already appeared in the literature as a factor of the spectral curve for $\mathrm{SU}(N)$ Chern--Simons theory  on the Seifert manifold $\Sigma_{2,3,3} \simeq S^3/\mathbb{T}_{24}$, where $\mathbb{T}_{24}$ is the binary tetrahedral group: the spectral parameter dependence of \eqref{eq:lambdatw} is indeed the same as that of \cite[Eq.(6.33)--(6.35)]{Borot:2015fxa}. This is large-$N$ dual to the topological A-string on the quotient $[\cO^{\oplus 2}_{\bbP^1}(-1)/\mathbb{T}_{24}]$ of the resolved conifold, which engineers the pure gauge theory with $\cG=E_6$. The family of spectral curves for the twisted $G_2$ Toda chain curve then sit as a three-parameter subfamily of the $E_6$ five-dimensional SW curves.
It is furthermore immediate to see that it returns the affine $G_2^\vee$ periodic Toda spectral curve  of  \cite{Martinec:1995by} in the $R_5 \to 0$ limit. \\


The quasi-polynomiality condition and the condition of existence of single-log solutions at large complex structure
uniquely fixes the privileged coordinate $t$-frame as
\begin{equation}
t_0=\log u_0,\;\;\;\;t_1=u_0^{\frac12} \left(u_1+2\right),\;\;\;\;t_2= u_0\left(\frac12 u_1^2+7  u_1+7 + u_2\right),
\end{equation}
in terms of which the structure constants read 
\begin{equation}
\mathsf{C}_{0i}^j={\footnotesize 
\left(
\begin{array}{ccc}
 \frac{6 t_2+18 \re^{t_0}-t_1^2+18 t_1 \re^{\frac{t_0}{2}}}{4} & \frac{2 t_2 t_1+12
   \re^{\frac{t_0}{2}} \left(4 t_1^2+t_2\right)-3 t_1^3-54 \re^{t_0} t_1}{8}  & \frac{4 t_2 t_1^2+6 \re^{\frac{t_0}{2}} \left(9 t_1^2+14 t_2\right) t_1-4 t_2^2+18
   \re^{t_0} \left(3 t_1^2-4 t_2-3 t_1^4\right)}{8}  \\
 -3 \re^{\frac{t_0}{2}} & -t_1^2+\frac{9}{2} \re^{\frac{t_0}{2}} t_1+t_2 & \re^{\frac{t_0}{2}} \left(9 t_1^2+6 t_2\right)+\frac{1}{4} \left(2 t_1 t_2-3 t_1^3\right) \\
 1 & 0 & 0 \\
\end{array}
\right)_i^j},
\end{equation}
\begin{equation}
\mathsf{C}_{1i}^j={\footnotesize 
\left(
\begin{array}{ccc}
 -3 \re^{\frac{t_0}{2}} & -t_1^2+\frac{9}{2} \re^{\frac{t_0}{2}} t_1+t_2 & \re^{\frac{t_0}{2}} \left(9 t_1^2+6 t_2\right)+\frac{1}{4} \left(2 t_1 t_2-3 t_1^3\right) \\
 2 & -3 t_1 & -\frac{3 t_1^2}{2}+9 \re^{\frac{t_0}{2}} t_1-t_2 \\
 0 & 1 & 0 \\
\end{array}
\right)_i^j,
}
\end{equation}
and $\mathsf{C}_{2i}^j=\delta_i^j$. \medskip

Unlike the twisted $B_2$ case, the spectral curve is not hyperelliptic here and it is more difficult to come up with a partial differential equation in the moduli which is already satisfied at the level of differentials. 
To fix the prepotential uniquely, we shall first  consider the full solution space of \eqref{eq:PF5prodgen}, without imposing any quasi-homogeneity condition of the form \eqref{eq:PF5Eulgen}. 
We find that the solutions thus obtained possess the following properties in the $z$-chart:
\begin{itemize}
\item all Taylor coefficients of $\cO(z_0^0)$ in the expansion around $z=0$ are unambiguously fixed only by \eqref{eq:PF5prodgen},
\item order by order in $z_0$, each solution comes with finitely many constant parameters that cannot be fixed purely by \eqref{eq:PF5prodz} due to the lack of the quasi-homogeneity equation \eqref{eq:PF5Eulz}, but imposing the existence of the prepotential in the form \eqref{eq:Fconstraint} \emph{uniquely} fixes such constant parameters.
\end{itemize}

Exactly as for the previous case of $\cG=B_2$, the perturbative prepotential contains as expected a sum of trilogarithms indexed by positive roots, but the long roots contributions are weighted by a relative factor of $4/(\alpha, \alpha)^2=1/9$ compared to the short ones, in disagreement with the gauge theory calculation already at the perturbative level (for $R_5\neq 0$). The higher order corrections in $z_0=q_0$ are computed efficiently and in a similar manner: rather unsurprisingly at this point, they also differ from the blow-up formula calculation of \cite{Keller:2012da}. We find
\bea  
-4\pi^3 \ri
\cF_{G_2} &=& 
\frac{4}{3} \log ^3q_1+\frac{1}{2} \log q_0 \log ^2q_1+2 \log ^2q_1 \log q_2 +\frac{7}{6}  \log q_1\log ^2q_2+\frac{7}{27} \log ^3q_2 \nn \\ &+& \frac{1}{6}
   \log q_0 \log ^2q_2+\frac{1}{2} \log q_0 \log q_2 \log q_1 \nn \\
   &+& q_1+\frac{q_2}{9}+\frac{q_1^2}{8}+q_2 q_1+\frac{q_2^2}{72}+\frac{q_1^3}{27}+
   q_1^2q_2+\frac{q_2^3}{243}+\frac{q_1^4}{64}+\frac{1}{9} q_1^3q_2+\frac{1}{8}  q_1^2q_2^2+\frac{q_2^4}{576}+\frac{q_1^5}{125}+\dots  \nn \\ &+& q_0 \bigg(q_1^4 q_2^2+2 q_1^5 q_2^2+3 q_1^6q_2^2 +2  q_1^5q_2^3+4  q_1^7q_2^2+6  q_1^6q_2^3+5  q_1^8q_2^2 +10  q_1^7q_2^3 +3  q_1^6q_2^4+\dots\bigg)
    \nn \\ &+&
    q_0^2 \bigg(\frac{1}{8} q_1^8 q_2^4+3 q_1^9 q_2^4+\frac{65}{4}  q_1^{10}q_2^4+3  q_1^9q_2^5 +55  q_1^{11}q_2^4 +25  q_1^{10}q_2^5+\dots \bigg) +\cO(q_0^3).
\eea

\section{Conclusion}
\label{sec:conclusion}
In this paper we formulated an algebraic construction of Picard--Fuchs D-modules for the periods of a class of B-model geometries related to $\cN=1$ super Yang--Mills theory on $\bbR^4 \times S^1$, including non-unitary gauge groups/non-toric engineering geometries. Our proposal passes a host of fine checks involving explicit period integral calculations on the B-model, and/or indirect tests from multi-instanton calculations in the gauge theory/A-model. One first upshot of our study is a direct confirmation of the validity of various proposals in the literature arising from  the theory of relativistic integrable systems (for simply-laced gauge groups) and from 5-brane web constructions (for classical groups): in doing so we also resolve a tension between two inequivalent proposals for symplectic gauge groups in \cite{Brandhuber:1997ua} and \cite{Li:2021rqr}, in favour of the latter.
A second surprising consequence is a disproof of the relevance of the twisted version of the affine relativistic Toda chain for non simply-laced gauge symmetry: the corresponding spectral curves are shown {\it not} to reproduce the low-energy prepotential of the gauge theory for BCFG types, in sharp contrast with the well-established situation for pure $\cN=2$ super Yang--Mills theory on $\bbR^4$.
\medskip 

Our results furthermore open several further avenues of investigation, which we  defer to future work. We describe some of these below.


\bit 

\item One obvious strand of implications is for the possible generalisation to a non-trivial $\Omega$-background, possibly including surface operators. In particular, in the Nekrasov--Shatashvili limit \cite{Nekrasov:2009rc}, it would be fascinating to explore to what extent the commutative algebra construction of \cref{sec:bcfg} of Picard--Fuchs systems from the ring of regular functions on the discriminant admits a canonical non-commutative quantisation: this aligns well with both the spirit of the quantum curve constructions of \cite{Aganagic:2011mi} and  the fact that the NS twisted effective superpotential is known to be computable from an $\hbar$-deformation of the usual special geometry relations \cite{Mironov:2009uv, Mironov:2009dv}. While it is clear how to promote the Seiberg--Witten curve to a quantum finite-difference operator, the notion of a non-commutative discriminant ideal relevant for the constructions of \cref{sec:bcfg} is less obvious. Moreover the fact that the system \eqref{eq:PF5prodgen} is holonomic rests on the underlying ring structure on the discriminant being commutative, which guarantees that the structure constants $\mathsf{C}_{ij}^k$ can be viewed as Christoffel symbols of a torsion-free connection on the Coulomb branch: how this feature would persist in the noncommutative setting would need to be clarified.
 
\item As far as the self-dual slice of the $\Omega$-background is concerned, the construction proposed here gives an effective technique for finding the mirror maps in the context of local mirror symmetry, and it would be interesting to combine it with the Chekhov--Eynard--Orantin topological recursion \cite{Chekhov:2005rr,Eynard:2007kz} to compute open and closed higher genus A-model amplitudes in the self-dual $\Omega$ background, as considered in
\cite{Alday:2009fs,Brini:2010fc,Awata:2010bz,Kozcaz:2010af} in relation to the AGT correspondence with surface operators, and studied in detail in \cite{Aganagic:2002wv,Brini:2008ik,Borot:2014kda,Borot:2015fxa,Brini:2017gfi} for ADE gauge groups in relation to the large $N$ duality with Chern--Simons theory on Seifert manifolds and related matrix models.  A fascinating relating development is the appearance of recursive structures for the self-dual theory arising from the non-autonomous Toda chain in four-dimensions, as recently been studied in \cite{Bonelli:2021rrg}, raising the prospects of a possible uplift to the 5d setup considered here.

\item Our story took place wholly in five dimensions and without hypermultiplets. The absence of matter leads to a symmetry under the involution $\lambda \leftrightarrow q_0/\lambda$ which simplifies the task of determining the PF ideal to calculations on the reduced (perturbative) Seiberg--Witten curves $\{\widetilde{\mathsf{P}}_{\cG;u}=0\}$, as in \cite{Ito:1997ur}. The lack of such symmetry may render the calculations more difficult in the case with non-trivial matter content, but it should possible to extend at least the treatment of \cref{sec:bcfg} to these cases as well. Furthermore, the trigonometric Frobenius manifolds of \cref{sec:ade} have canonical elliptic deformations in terms of Frobenius structures on orbits of Jacobi groups, as in \cite{MR1775220,MR1797874}: it is only natural to conjecture that these would play for a 6D uplift of the theory the same role that their rational (Weyl) and trigonometric (extended affine Weyl) had for the four and five-dimensional setup respectively. Note that for $\cG=\mathrm{SU}(N)$ their almost-dual Frobenius manifolds \cite{MR2070050} are indeed already known to closely reproduce the six-dimensional perturbative gauge theory prepotential \cite{Marshakov:1997cj} in terms of Beilinson's elliptic polylogarithm, in exactly the same way as the almost-duals of their rational \cite{MR2070050} and trigonometric \cite{Stedman:2021wqs} limits Frobenius manifolds gave the 1-loop prepotential of the four and five-dimensional gauge theories respectively.

\item The surprising failure of the twisted affine relativistic Toda system to reproduce the low energy effective action of the theory with BCFG gauge symmetry raises several interesting questions. The first, obvious one is what would then be the correct integrable model underlying the gauge theory. This should be a relativistic deformation of the twisted affine Toda chain which is qualitatively different from the natural one considered here by Dynkin folding. That would be similar to the way in which the integrable models underlying the $\mathrm{SU}(N)$ theory at different Chern--Simons levels provide inequivalent relativistic deformations of the periodic Toda chain, to which they however all reduce in the four-dimensional/non-relativistic limit \cite{Brini:2008rh,Eager:2011dp}. For classical groups, the dimer model strategy of \cite{Eager:2011dp} applied to the M-theory curves of \cite{Brandhuber:1997ua,Li:2021rqr} might be a promising avenue to make headway into this, and possibly hint to a solution encompassing exceptional non-simply-laced gauge groups as well.

\item The reverse question is also of some non-trivial relevance, i.e. what is the physical and/or geometrical significance of the B-model on the local geometry associated to the spectral curves of the twisted affine relativistic Toda chain? Part of the answer is already known: 
these curves indeed arise via affine Dynkin folding on a sublocus of the K\"ahler moduli space of the topological A-model on ADE-orbifolds of the resolved conifold, where they have already been found {\it ante litteram} to describe (at small 't Hooft coupling/small K\"ahler volume) the large $N$ limit of Chern--Simons theory on Seifert manifolds of positive curvature \cite{Borot:2015fxa}. So even though our results in \cref{sec:PFToda} disprove that the resulting special geometry prepotentials are the gauge theory prepotentials of the pure KK theory with $\cG=\mathrm{BCFG}$, there is still however room for a gauge theory angle on them. For example, although the $G_2$ twisted relativistic Toda curves do not describe the weak coupling expansion of the pure $G_2$ theory in 5d, they do sit as the Seiberg--Witten curves of the $E_6$ theory on a special 3-dimensional locus of its Coulomb branch \cite{Borot:2015fxa} determined by the folding procedure. It would nonetheless be interesting to give a more poignant characterisation of the large radius expansions of \cref{sec:PFToda}, whether intrinsically or in terms of a suitable deformation of the $\cN=2$ KK theory with non-simply-laced gauge symmetry.

\eit 



\bibliography{miabiblio}

\end{document}